\documentclass[11pt]{article}
\pdfoutput=1
\usepackage{amsmath,amssymb,epsfig,amsfonts}
\usepackage{graphicx}
\usepackage{subfig}
\usepackage[usenames, dvipsnames]{color}
\usepackage{cite}
\usepackage{multirow}
\usepackage{hyperref}
\usepackage{verbatim} 
\usepackage{enumitem}
\usepackage{rotating}
\usepackage{xcolor}
\usepackage{multirow}
\usepackage{bm}
\usepackage[normalem]{ulem}
\usepackage{cleveref}
\usepackage{slashed}
\usepackage{lscape}

\usepackage{tikz}
\usetikzlibrary{positioning}
\usetikzlibrary{arrows}

\graphicspath{ {figs/} }

\makeatletter

\newcommand{\lb}{\left(}
\newcommand{\rb}{\right)}
\newcommand{\lbb}{\left[}
\newcommand{\rbb}{\right]}
\newcommand{\lbbb}{\left\{}
\newcommand{\rbbb}{\right\}}
\newcommand{\tn}[1]{\textnormal{#1}}

\newcommand{\be}{\begin{equation}}
\newcommand{\ee}{\end{equation}}
\newcommand{\ba}{\begin{aligned}}
\newcommand{\ea}{\end{aligned}}

\newcommand{\C}{\mathbb{C}}
\renewcommand{\P}{\mathbb{P}}

\newcommand{\bea}{\begin{eqnarray}}
\newcommand{\eea}{\end{eqnarray}}

\def\unit{{1\kern-.65ex {\rm l}}}
\def\1{{1\kern-.65ex {\rm l}}}

\def\CF{{\cal F}}

\def\CN{{\cal N}}
\def\CO{{\cal O}}

\newcount\hour \newcount\minute
\hour=\time \divide \hour by 60
\minute=\time
\def\now{%
\ifnum \hour<13
  \ifnum \hour=0 \advance \hour by 12 \number\hour:\else \number\hour:\fi%
     \ifnum \minute<10 0\fi%
     \number\minute%
\ A.M.%
\else \advance \hour by -12 \number\hour:%
  \ifnum \minute<10 0\fi%
  \number\minute%
  \ P.M.%
\fi%
}

\makeatother

\begin{document}

\baselineskip=18pt  
\numberwithin{equation}{section}  
\allowdisplaybreaks  


%
%


\thispagestyle{empty}


\vspace*{1cm} 
\begin{center}
{\Huge {5d SCFTs from $(E_n,E_m)$ Conformal Matter}
}

\vspace*{1.5cm}
Max H\"ubner\\
\vspace*{0.5cm} 
{\it Mathematical Institute, University of Oxford, \\
Andrew-Wiles Building,  Woodstock Road, Oxford, OX2 6GG, UK}\\
\vspace*{1cm}
\end{center}
\noindent We determine 5d $\CN=1$ SCFTs originating from 6d $(E_n,E_m)$ conformal matter theories with $n\neq m$ by circle reduction and mass deformations. The marginal geometries are constructed and we derive their combined fiber diagrams (CFDs). The CFDs allow for an enumeration of descendant SCFTs obtained by decoupling matter hypermultiplets and a description of candidate weakly coupled quivers. 

%
\newpage

\tableofcontents




\section{Introduction}

5d $\CN=1$ superconformal field theories (SCFTs) are non-perturbative in nature. The first such theories were understood as the UV completion of non-renormal\-izable 5d supersymmetric gauge theories realized by an embedding into string theory \cite{Seiberg:1996bd}. Geometrizations of these constructions have proven to be an efficient method for systemizing large classes of 5d SCFTs \cite{Intriligator:1997pq, Morrison:1996xf, DelZotto:2017pti, Xie:2017pfl, Tian:2018icz, Bhardwaj:2018vuu, Bhardwaj:2018yhy, Jefferson:2018irk, Apruzzi:2018nre, Closset:2018bjz, Apruzzi:2019vpe, Apruzzi:2019opn, Apruzzi:2019enx, Apruzzi:2019kgb, Bhardwaj:2019ngx,Bhardwaj:2019fzv, Saxena:2019wuy, Closset:2019juk, Bhardwaj:2020gyu, Corvilain:2020tfb, Morrison:2020ool, Albertini:2020mdx} and have greatly expanded the list of known examples and their properties.

The classification of 6d $\CN=(1,0)$ SCFTs is organized by an enumeration of admissible F-theory geometries \cite{Heckman:2013pva,Heckman:2015bfa}. These geometries are elliptically fibered Calabi-Yau 3-folds with the classification amounting to a characterization of the possible base geometries and permitted singular fibers above these. The base geometries are built from non-Higgsable clusters connected by conformal matter theories and other links. These conformal matter theories \cite{DelZotto:2014hpa} are SCFTs themselves and essential building blocks of the classification result. Sharpening this classification, it was argued in \cite{Heckman:2018pqx} that under certain Higgs and tensor branch flows all classified 6d $\CN=(1,0)$ SCFTs are generated starting from a small set of UV-progenitor theories, the rank $k$ $(E_8,G_{\tn{ADE}})$ orbi-instanton theories.

Circle compactifications of 6d $\CN=(1,0)$ SCFTs yield 5d KK-theories for which suitable mass deformations trigger an RG flow to 5d $\CN=1$ SCFTs. Geometrically the mass deformations are realized as partial resolutions of the fiber singularities of the Calabi-Yau 3-fold used to engineer the 6d SCFT. By M-/F-theory duality the 5d SCFT is then realized by M-theory on this partially resolved Calabi-Yau 3-fold. This observation has been systematized to a classification programme in \cite{Jefferson:2018irk, Bhardwaj:2018yhy, Bhardwaj:2018vuu, Apruzzi:2019opn, Apruzzi:2019enx, Apruzzi:2019kgb, Bhardwaj:2019fzv} and suggests an avenue to utilize the classification of 6d $\CN=(1,0)$ SCFTs for 5d $\CN=1$ SCFTs. As a first step in this programmme 5d SCFTs originating from 6d conformal matter theories of types $(E_n,A_m),(D_n,D_n),(E_n,E_n)$ and non-Higgsable clusters were analysed \cite{Apruzzi:2019opn, Apruzzi:2019enx, Apruzzi:2019kgb}. We report on 5d SCFTs originating from the 6d $(E_n,E_m)$ with $n\neq m$ conformal matter theories. This serves as an initial step in analyzing the associated circle reductions of the 6d UV-progenitor theories with the ultimate goal of systematically determining all 5d theories that descend through 6d SCFTs. Extending this approach to include 5d SCFTs reached through RG flows triggered by Higgs branch vacuum expectation values would further connect to recent results in \cite{Ferlito:2017xdq, Cabrera:2018jxt, Cabrera:2018ann, Cabrera:2018uvz, Bourget:2019rtl, Bourget:2019aer, Eckhard:2020jyr, Bourget:2020gzi}. 

We study the 5d $\CN=1$ SCFTs at generic points of their Couloumb branches via the associated resolved Calabi-Yau 3-folds. These exhibit a non-flat fiber given by a reducible surface $S=\cup_{k}S_k$ which collapses in the singular limit and characterizes the SCFT. Much of the resolution independent data of the surface $S$ can be subsumed into combined fiber diagrams (CFDs), introduced in \cite{Apruzzi:2019vpe, Apruzzi:2019opn}, which then manifestly encode many properties of the SCFT such as the superconformal flavor symmetry, BPS states, mass deformations and possible quiver descriptions. This constitutes a uniform geometric formulation of many known results from field  theoretic and brane web considerations \cite{Aharony:1997ju, Kim:2012gu, Tachikawa:2015mha, Zafrir:2014ywa, Bergman:2013aca, Mitev:2014jza, Yonekura:2015ksa, Hwang:2014uwa, Zafrir:2015uaa, Hayashi:2018lyv, Uhlemann:2019ypp}.

This paper is organized as follows. In section \ref{sec:MarginalGeometries} we discuss the singular geometries realizing  $(E_n,E_m)$ conformal matter and their resolutions. We compute the marginal fiber diagrams and the underlying geometries of the irreducible components of the non-flat surface $S$. We conclude the section with a derivation of the marginal CFDs for $(E_n,E_m)$ conformal matter. In section \ref{sec:DescendantsAndQuivers} we utilize the CFDs to enumerate 5d SCFTs descending from the marginal theories via mass deformations. Furthermore we derive possible weakly coupled quiver descriptions of the marginal theories and its descendants, which are not excluded by consistency constraints imposed by their CFDs and those derived in \cite{Jefferson:2017ahm} .

\section{Marginal Geometries for Conformal Matter}
\label{sec:MarginalGeometries}

Marginal theories in 5d are circle reductions of 6d $\CN=(1,0)$ gauge theories which UV complete to a 6d $\CN=(1,0)$ SCFT. Consider a marginal theory given by a 5d $\CN=1$ gauge theory with gauge group $G$ and gauge algebra $\mathfrak{g}$ of rank $r=\tn{rank}\,\mathfrak{g}$ coupled to massive matter. At generic points of its Coulomb branch the field content is given by $r$ massless $U(1)$ vector multiplets and massive hypermultiplets associated with the W-bosons of the broken gauge symmetry and the original matter multiplets. Integrating out the W-bosons the dynamics of the low energy effective theory is governed by a prepotential $\CF$ cubic in the $U(1)$ vector multiplets. The terms involving the scalars $\phi^i$ of the $r$ vector multiplets parametrizing the Coulomb branch reads
\be\label{eq:PrePot}
\CF=\left(\frac{1}{2 g_{5d}^{2}} h_{i j} \phi^{i} \phi^{j}+\frac{k}{6} d_{ijk} \phi^{i} \phi^{j} \phi^{k}\right)+\frac{1}{12}\left(\sum_{\alpha \in \Phi_{\mathfrak{g}}}\left|\alpha_{i} \phi^{i}\right|^{3}-\sum_{\mathbf{R}_{f}} \sum_{\lambda \in \mathbf{W}_{\mathbf{R}_{f}}}\left|\lambda_{i} \phi^{i}+m_{f}\right|^{3}\right)\,.
\ee
Here $g_{5d}$ is the 5d Yang-Mills coupling constant, $h_{i j}=\tn{tr}\, T_iT_j$ is the metric on the moduli space involving the Lie algebra generators $T_i$, the integer $k$ is the Chern-Simons level, $d_{ijk}=\frac{1}{2} \operatorname{tr}\left(T_{i}\left(T_{j} T_{k}+T_{k} T_{j}\right)\right)$ is a symmetric group theoretic quantity, $\mathbf{R}_{f}$ are the representations of the massive hypermultiplets with masses $m_f$ and $\mathbf{W}_{\mathbf{R}_{f}}$ is the weight system of these representations \cite{Seiberg:1996bd,Intriligator:1997pq}.

Let the 6d $\CN=(1,0)$ SCFT associated to this marginal theory be realized by F-theory on a singular Calabi-Yau 3-fold $X_3$. Then the marginal theory is realized by M-theory at low energies on a crepant resolution $Z_3$ of this Calabi-Yau 3-fold. Given a basis of K\"ahler classes $J_i\in H^{1,1}(Z_3)$ of unit volume and defining coordinates $J=\phi^iJ_i$ the triple intersections $c_{ijk}=S_i\cdot S_j\cdot S_k=\int_{Z_3}J_i\wedge J_j\wedge J_k$ set the cubic term of the prepotential $ \frac{1}{6}c_{ijk}\phi^i\phi^j\phi^k \subset \CF$ describing the marginal theory \cite{Ferrara:1996hh, Intriligator:1997pq}. Here $S_i$ are complex surfaces dual to the basis $J_i$ of K\"ahler classes. Masses are set by volumes of two-cycles within $Z_3$ and their intersection structure determines the representation theoretic details of \eqref{eq:PrePot}.  
 
Mass deformations of the marginal theory and subsequent RG flows of the 5d theories correspond in geometry to partial singular limits $Z_3\rightarrow X_3'$ parametrized by the RG flow. For suitably chosen mass deformations this procedure results in a family of 5d SCFTs enumerated by partial resolutions of $X_3$. These SCFTs are thereby derived from the smooth geometry $Z_3$ of the 5d marginal theory, they are referred to as descendants of the marginal theory. Consequently the starting point to the analysis of this tree of descendants and their properties is the marginal geometry $Z_3$.

\subsection{Singular Elliptically Fibered Calabi-Yau 3-folds}

\begin{figure}
  \centering
  \includegraphics[width=13cm]{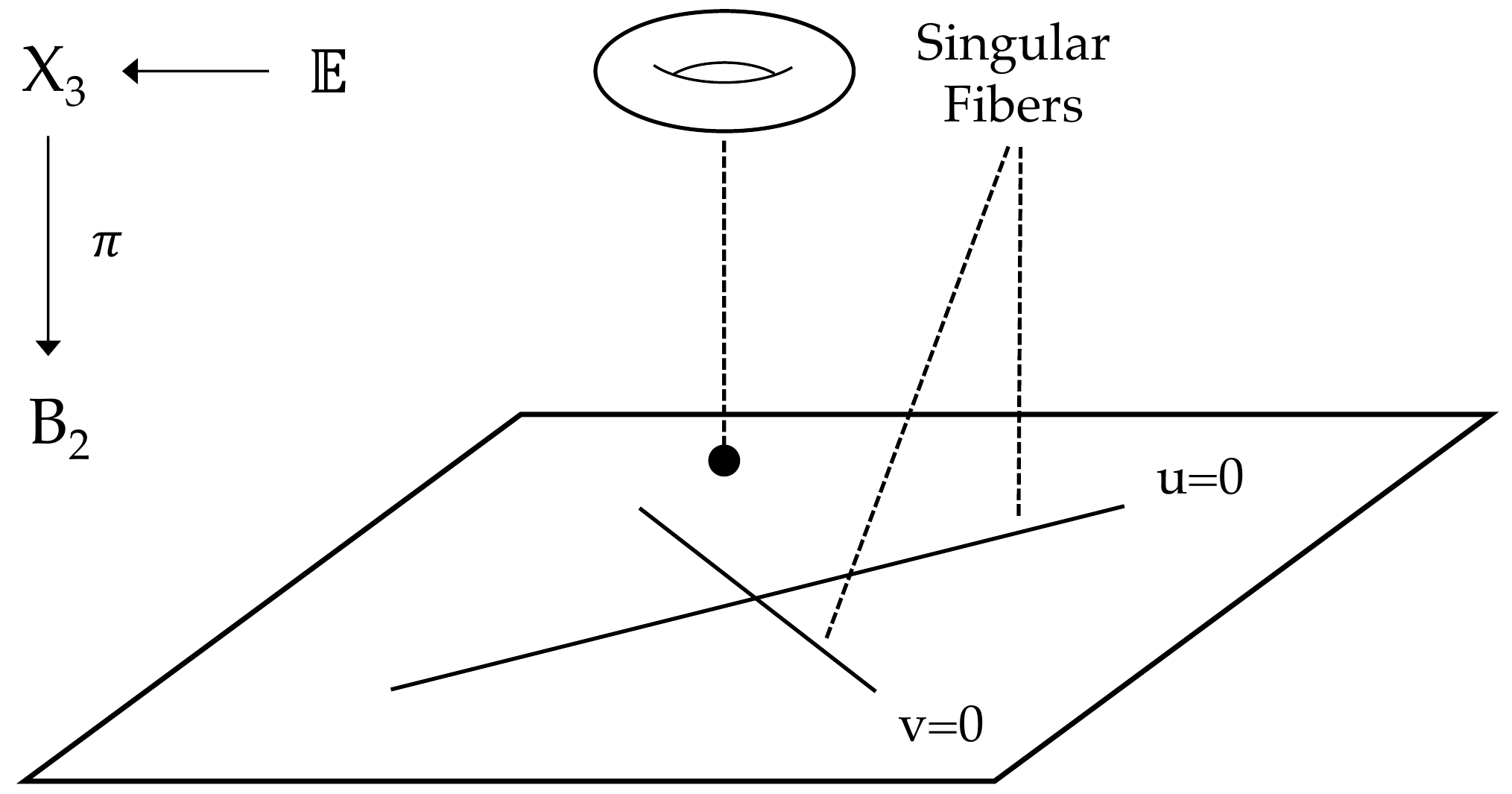}
    \caption{Sketch of the singular Calabi-Yau 3-fold geometry $X_3$ \eqref{eq:GeneralTateForm}. Minimal singularities of Kodaira-type are supported along base divisors $u,v=0$ and enhance to a non-minimal singularity upon collision at the origin.}\label{fig:SingGeo}
\end{figure} 

The geometries we use to engineer 5d $\CN=1$ conformal matter are non-compact singular elliptically fibered Calabi-Yau 3-folds with a holomorphic section and non-isolated, non-minimal singularities. We begin the construction of these geometries with an elliptically fibered 3-fold $X_3$ and write
\be
\mathbb{E}\hookrightarrow X_3\xrightarrow[]{\pi} B_2\,,
\ee 
where $\mathbb{E}$ is the elliptic fiber and $B_2$ is the 2-fold K\"ahler base. The 3-fold $X_3$ is realized as a hypersurface within the weighted projective bundle over $B_2$ given by
\be\label{eq:AmbientProjectiveBundle}
X_4=\P^{231}\left( K^{-2}_{B_2}\oplus K^{-3}_{B_2}\oplus \mathcal{O}\right)\,,
\ee
where $K_{B_2}$ denotes the canonical bundle of the base $B_2$. Let $\lb u,v\rb$ be global complex coordinates on the base $B_2$ and take $[x:y:w]$ to be the homogenous coordinates of the fiber. The elliptic fiber is then realized by the Tate form \cite{Tate,Bershadsky:1996nh,Katz:2011qp}
\be\ba\label{eq:GeneralTateForm2}
X_3\,:\qquad 0&=y^2-x^3+a_1 xyw-a_2x^2w^2 +a_3y w^3-a_4xw^4-a_6w^6\,,
\ea\ee
with the sections $a_n\in \Gamma(B_2,K_{B_2}^{-n})$ encoding base dependence. The holomorphic section of the fibration is given by $s:B_2\rightarrow X_3$ mapping as $(u,v)\mapsto [1:1:0]$. The divisor $\sigma=\lbb w=0\rbb\subset X_4$ thus intersects $X_3$ precisely along the image of $s$ and once in each fiber. 

We introduce singularities of Kodaira type above the base divisors $S_u=\{u=0\}\subset B_2$ and $S_v=\{v=0\}\subset B_2$. This amounts to prescribing the vanishing orders of the sections $a_n\in \Gamma(B_2,K_{B_2}^{-n})$ along $S_u,S_v$ when expanded in the coordinates $(u,v)$. Denoting the two sets of exponents for $u$ and $v$ by $(i_{1},i_{2},i_{3},i_{4},i_{6})$ and $(j_{1},j_{2},j_{3},j_{4},j_{6})$ respectively the Tate form \eqref{eq:GeneralTateForm2} now reads
\be\ba\label{eq:GeneralTateForm}
X_3\,:\qquad 0&=y^2-x^3+b_1 xyw \left( u^{i_{1}}v^{j_{1}}\right)-b_2x^2w^2\left( u^{i_{2}}v^{j_{2}}\right)\\ &~~~+b_3y w^3\left( u^{i_{3}}v^{j_{3}}\right)-b_4xw^4\left( u^{i_{4}}v^{j_{4}}\right)-b_6w^6\left( u^{i_{6}}v^{j_{6}}\right)\,,
\ea\ee
with singularities along $[0:0:1]\in \mathbb{E}$ over $S_u,S_v$ and a generically non-minimal singularity at the point $u=v=0$ in the base. We sketch the setup in figure \ref{fig:SingGeo}. The classes of various section are
\be\ba
&x~:~2\sigma +2c_1\,,\qquad && y~:~3\sigma +3c_1\,,\\
&w~:~\sigma\,, \qquad && b_n~:~nc_1-i_{n}S_u-j_{n}S_v \,, \\
&u~:~S_u\,,\qquad && v~:~S_v\,,
\ea\ee
where $c_1=c_1\left(TB_2\right)$ is the first chern class of the base $B_2$. 

\subsection{Resolution of Singularities}

We resolve the singularities of the Calabi-Yau 3-fold \eqref{eq:GeneralTateForm} by blowing up in the base once to remove the non-minimal singularity located at $u=v=0$ and subsequently resolving the codimension 1 and 2 singularities in the fiber. We adhere to the resolution procedure and notation presented in \cite{Lawrie:2012gg, Hayashi:2014kca}, which we reintroduce where necessary.

The non-minimal singularity is removed by the blowup
\be\label{eq:BaseBlowUp}
u\rightarrow \epsilon u\,, \quad v\rightarrow \epsilon v\,,
\ee
together with the rescaling $x\rightarrow \epsilon^2 x$ and $y\rightarrow \epsilon^3 y $ which introduces the exceptional divisor $E=\left\{\epsilon=0\right\}$ in the base $\widetilde B_2\rightarrow B_2$. The rescaling of the two sections $x,y$ previously belonging to $\Gamma( B_2,K_{B_2}^{-k})$ with $k=2,3$ respectively is due to the canonical bundle shifting to $K_{\widetilde B_2}=K_{B_2}+E$. Physically this resolution amounts to moving onto the tensor branch of the SCFT. The chern class $c_1$ and base divisor classes $S_u,S_v$ are all shifted by a copy of the exceptional divisor $E$ to
\begin{alignat*}{3}
x\,:&~2\sigma +2c_1-2E\,,\qquad && y~&&:~3\sigma +3c_1-3E\,,\\
w~:&~\sigma\,, \qquad && b_n\,&&:~nc_1-nE-i_{n}S_u-j_{n}S_v+i_{n}E+j_{n}E \,, \\
u\,:&~S_u-E\,,\qquad && v\,&&:~S_v-E\,.
\end{alignat*}
The partially resolved geometry $Y_3$ is explicitly given by substituting \eqref{eq:BaseBlowUp} into \eqref{eq:GeneralTateForm} with an overall power $\epsilon^6$ removed by a proper transform
\be\ba\label{eq:GeneralTateFormBB}
Y_3\,:\quad 0&=y^2-x^3+b_1 xyw \left( u^{i_{1}}v^{j_{1}}\right)\epsilon^{i_{1}+j_{1}-1}-b_2x^2w^2\left( u^{i_{2}}v^{j_{2}}\right)\epsilon^{i_{2}+j_{2}-2}\\ &~~~+b_3y w^3\left( u^{i_{3}}v^{j_{3}}\right)\epsilon^{i_{3}+j_{3}-3}-b_4xw^4\left( u^{i_{4}}v^{j_{4}}\right)\epsilon^{i_{4}+j_{4}-4}-b_6w^6\left( u^{i_{6}}v^{j_{6}}\right)\epsilon^{i_{6}+j_{6}-6}\,.
\ea\ee
This gives a hypersurface in $Y_4=\P\left(\mathcal{O}\oplus K^{-2}_{\widetilde B_2}\oplus K^{-3}_{\widetilde B_2}\right)$ where $\widetilde B_2$ is the blowup of the base $B_2$. The coordinates $u,v$ can no longer vanish simultanesouly, the non-minimal singularity is removed.

The Calabi-Yau 3-fold $Y_3$ still exhibits singularities in codimension 1 and 2 which can be removed with additional blowups in the ambient space \cite{Marsano:2009gv, Marsano:2011hv, Krause:2011xj, Esole:2017kyr}. Singularities at
\be\ba\label{eq:SingLocus}
\tn{Codim 1 Singularities}\,:&\qquad 0=s_a=s_b=s_c\,, \\
\tn{Codim 2 Singularities}\,:&\qquad 0=s_i=s_j\,, 
\ea\ee
where $s_a,s_i$ are place holders for generic sections of the Calabi-Yau, are resolved by the replacements 
\be\ba\label{eq:BlowUps2}
\tn{Codim 1 Blowup}\,:&\qquad s_a\rightarrow s_as_d\,,\,  &&s_b\rightarrow s_bs_d\,, \quad s_c\rightarrow s_cs_d\,, \\
\tn{Codim 2 Blowup}\,:&\qquad s_i\rightarrow s_is_k\,, &&s_j\rightarrow s_js_k\,, 
\ea\ee
together with a proper transform which removes a factor of $s_d^2,s_k$ from the transformed Tate form. We abbreviated these replacements together with their proper transforms by
\be\ba\label{eq:BlowUps}
\tn{Codim 1 Blowup}\,:&\qquad (s_a,s_b,s_c;s_d)\,, \\
\tn{Codim 2 Blowup}\,:&\qquad (s_i,s_j;s_k)\,. 
\ea\ee
These blowups introduce the exceptional divisors $D_{s_d}, D_{s_k}\subset \widetilde Y_4$ in the blowup of $Y_4$. We iterate these resolutions and obtain a smooth Calabi-Yau 3-fold $Z_3$ realized as a hypersurface in $Z_4$ given by the multi blowup of $Y_4$
\be\label{eq:SmoothGeometry}
Z_3\subset Z_4\,:\qquad \widetilde{\mathbb{E}}\hookrightarrow Z_3 \xrightarrow[]{\pi}  \widetilde{B}_2\,.
\ee
This fibration is non-flat, i.e. it contains fibers $S$ of complex dimension 2 which encode the SCFT data. Non-flat fibrations of this kind have most recently been studied in \cite{Xie:2017pfl, Tian:2018icz, Apruzzi:2018nre}.
 
\subsection{Geometries for $(E_n,E_m)$ Conformal Matter }
\label{sec:EnEmGeometries}

We select two distinct sets of vanishing orders for the ordered set of coefficients $(b_1,b_2,b_3,b_4,b_6)$ appearing in \eqref{eq:GeneralTateFormBB} corresponding to E-type singularities
\be\ba\label{eq:VanishingOrders}
E_6\,:&\quad (1,2,2,3,5)\,, \\
E_7\,:&\quad (1,2,3,3,5)\,,\\
E_8\,:&\quad (1,2,3,4,5)\,,\\
\ea\ee
and substitute these into the Tate-model \eqref{eq:GeneralTateForm}. The vanishing orders \eqref{eq:VanishingOrders} can be found in the lists of possible elliptic fiber degeneracies presented e.g. in \cite{Bershadsky:1996nh,Lawrie:2012gg}. We blowup in the base $B_2$ as in \eqref{eq:BaseBlowUp} to find the Tate models for the partially resolved geometries
\be\ba\label{eq:SingTates}
(E_6,E_7)\,:&\quad 0=y^2-x^3+b_1uvxy\epsilon-b_2u^2v^2x^2\epsilon^2 +b_3u^3v^2y\epsilon^2 -b_4u^3v^3x\epsilon^2-b_6u^5v^5\epsilon^4\,, \\
(E_6,E_8)\,:&\quad 0=y^2-x^3+b_1uvxy\epsilon-b_2u^2v^2x^2\epsilon^2 + b_3u^3v^2y\epsilon^2 -b_4u^4v^3x\delta_1^3-b_6u^5v^5\epsilon^4  \,,\\
(E_7,E_8)\,:&\quad 0=y^2-x^3+b_1uvxy\epsilon-b_2u^2v^2x^2\epsilon^2 +b_3u^3v^3y\epsilon^3 -b_4u^3v^4x\epsilon^3-b_6u^5v^5\epsilon^4\,.\\
\ea\ee
We resolve each E-type singularity individually using two of the blowup sequences 
\be
\ba \label{eq:ResolutionSequenceMarginal}
E_6\,:&\quad
\{x, y, u; u_1\}, \{x, y, u_1; u_2\},\{y, u_1, u_2; u_3\}, \{y, u_1; u_4\}, \{y, u_2; 
u_5\}, \cr &\quad \{y, u_3; u_6\}, \{u_1, u_4; u_7\}, \{u_4, u_3; u_8\}, \cr 
E_7\,:&\quad
\{x, y, u; u_1\}, \{x, y, u_1; u_2\},\{y, u_1;  u_3\}, \{y, u_2; u_4\}, \{u_2, u_3; 
u_5\}, \cr &\quad \{u_1, u_3; u_6\}, \{u_2, u_4; u_7\}, \{u_3, u_4; u_8\}, \{u_4.u_5; u_9\},\{u_5,u_8; u_{10}\} \cr & \quad\{u_3,u_5; u_{11}\},\cr 
E_8\,:&\quad
\{x, y, u; u_1\}, \{x, y, u_1; u_2\}, \{y, u_2; u_3\}, \{y, u_1, u_3; u_4\}, \{y, u_1; u_5\},  \cr &\quad \{u_1, u_3; u_6\}, \{u_2, u_3; u_7\}, \{u_3, u_4; u_8\}, \{u_1, u_4; u_9\}, \{u_1, u_5; u_{10}\},  \cr &\quad \{u_3, u_6; u_{11}\}, \{u_4, u_6; u_{12}\}, \{u_6, u_8; u_{13}\}, \{u_3, u_{11}; u_{14}\}, \{u_{8}, u_{11}; u_{15}\}\,.
\ea
\ee
We have listed these blowups in the notation introduced in \eqref{eq:BlowUps} here with the generic sections $s_a,s_i$ now explicitly given by $x,y,u_i,v_i$. The Cartan divisors intersecting according to the affine $E_n$-Dynkin diagrams among the exceptional divisors of \eqref{eq:ResolutionSequenceMarginal} are 
\be\ba
E_6\,:&\quad\{u,u_8,u_6,u_7,u_5,u_3,u_2\}=\{D_{\alpha_i}^{E_6}\,|\,i=0,\dots,6 \}\,, \\
	E_7\,:&\quad\{u,u_{6},u_{11},u_{10},u_{9},u_{7},u_{4},u_8\}=\{D_{\alpha_i}^{E_7}\,|\,i=0,\dots,7 \}\,,  \\	
			E_8\,:&\quad\{u,u_{10},u_9,u_{12},u_{13},u_{15},u_{14},u_7,u_{8}\}=\{D_{\alpha_i}^{E_8}\,|\,i=0,\dots,8 \} \,,
\ea\ee
Finally the remaining singularities are resolved by the cross term blowups, i.e. blowups involving a mix of sections and introducing $\delta_k$, 
\be\ba\label{eq:blowups}
\lb E_6^{(v)},E_7^{(u)}\rb\,:&\quad \{\epsilon,u_4;\delta_2\},\{\epsilon,v_5;\delta_3\},\{\delta_2,u_4;\delta_4\},\{\delta_3,v_5;\delta_5\},\{x,y,\epsilon;\delta_6\}, \{y,\delta_6;\delta_7\},\\
&\quad\{\delta_6,\delta_7;\delta_8\}, \{x,\delta_7;\delta_9\},\{x,y,\delta_2;\delta_{10}\},\{x,y,\delta_3;\delta_{11}\}\,,\\
\lb E_6^{(v)},E_8^{(u)}\rb\,:&\quad  \{\epsilon,u_3;\delta_2\},\{\epsilon,v_5;\delta_3\},\{\delta_2,u_3;\delta_4\}, \{\delta_3,v_5;\delta_5\},\{\delta_4,u_3;\delta_6\},\{\delta_6,u_3;\delta_7\},  \\
					&\quad\{x,y,\epsilon;\delta_8\},\{y,\delta_8;\delta_9\},\{\delta_8,\delta_9;\delta_{10}\}, \{x,\delta_9;\delta_{11}\},\{\delta_9,\delta_{11};\delta_{12}\}, \\
					&\quad\{\delta_9,\delta_{12};\delta_{13}\},\{x,y,\delta_{2};\delta_{14}\},\{y,\delta_{14};\delta_{15}\},\{\delta_{14},\delta_{15};\delta_{16}\},
					\{x,y,\delta_3;\delta_{17} \},\\
				&\quad	\{y,\delta_{17};\delta_{18}\},\{x,y,\delta_4;\delta_{19}\},\{y,\delta_2,\delta_{10};\delta_{20}\},\{y,\delta_3,\delta_{10};\delta_{21}\}\,, \\
\lb E_7^{(u)},E_8^{(v)}\rb\,:&\quad	\{\epsilon,u_4;\delta_2\},\{\epsilon,v_3;\delta_3\},\{\delta_2,u_4;\delta_4\}, \{\delta_3,v_3;\delta_5\},\{\delta_4,u_4;\delta_6\},\{\delta_5,v_3;\delta_7\},  \\
			&\quad	\{v_3,\delta_7;\delta_8\},\{x,y,\epsilon;\delta_9\},\{y,\delta_9;\delta_{10}\}, \{\delta_9,\delta_{10};\delta_{11}\},\{x,\delta_{10};\delta_{12}\}, \\
			&\quad	\{\delta_{10},\delta_{12};\delta_{13}\},\{\delta_{10},\delta_{13};\delta_{14}\},\{x,y,\delta_{2};\delta_{15}\},\{y,\delta_{15};\delta_{16}\}, \{\delta_{15},\delta_{16};\delta_{17}\},\\
			&\quad	\{x,y,\delta_3;\delta_{18} \},\{y,\delta_{18};\delta_{19}\},\{\delta_{18},\delta_{19};\delta_{20}\},\{x,y,\delta_4;\delta_{21}\},\{x,y,\delta_5;\delta_{22}\},\\ &\quad \{y,\delta_2,\delta_{12};\delta_{23}\},\{y,\delta_3,\delta_{12};\delta_{24}\}\,,
\ea\ee
where the superscript $\alpha$ in $E_n^{(\alpha)}$ denotes the choice of base coordinate over which the $E_n$ singularity is fibered in the Tate model \eqref{eq:GeneralTateFormBB}.

The projective relations introduced by the blowups prohibit the sections $\delta_2$ and $\delta_2,\delta_9,\delta_{14}$ and $\delta_2,\delta_{10},\delta_{15},\delta_{18}$ from vanishing for $(E_6,E_7)$ and $(E_6,E_8)$ and $(E_7,E_8)$ respectively. For all other sections $\epsilon,\delta_i$ restricting the associated divisors to the smooth Calabi-Yau 3-fold yields an irreducible complex surface. The number of these surface components is the rank $r$ of the associated SCFT, we have
\be\label{eq:Ranks}
(E_6,E_7)\,: r=10\,, \qquad (E_6,E_8)\,: r=18\,, \qquad (E_7,E_8)\,: r=20\,.
\ee

\subsection{Intersection Ring}

The intersection ring of the divisors of the fully resolved Calabi-Yau 3-fold $Z_3\subset Z_4$ determines the 5d physics. There are vertical and horizontal divisors in $Z_3$. The vertical divisors are pull backs of divisors in the base $\widetilde B_2$, i.e.  $\pi^{*}(E), \pi^{*}(S_u), \pi^{*}(S_v)$, while the horizontal divisors are the exceptional divisors $D_{u_i},D_{v_i},D_{\delta_i}\subset Z_4$ introduced in the blowups of \eqref{eq:BlowUps} restricted to $Z_3$ together with the divisor associated to the holomorphic section $\sigma$.

The intersection rings of the 3-fold $Z_3$ and its base $\widetilde B_2$ are related as
\be\ba\label{eq:Intersections}
\sigma\cdot |_{Z_3\:}\sigma \cdot|_{Z_3\:}  \pi^{*}(V)&=-c_1(T\widetilde B_2)\cdot|_{\widetilde B_2~}V\,,\\
\sigma\cdot |_{Z_3\:}  \pi^{*}(V) \cdot |_{Z_3\:}  \pi^{*}(V) &=V\cdot|_{\widetilde B_2\:} V \,,\\
V \cdot |_{Z_3\:} \pi^{*}(V) \cdot |_{Z_3\:}  \pi^{*}(V)&=0\,,\\
\ea\ee
where $V=E,S_u,S_v$. The first chern class of the base $B_2=\{(u,v)\}=\C^2$ prior to the blowup vanishes and consequently that of the blowup $\widetilde B_2$ evaluates to $c_1 (T\widetilde B_2)=c_1(\bar K_{\widetilde B_2})=-c_1( K_{\widetilde B_2})=-E$. After the base blowup \eqref{eq:BaseBlowUp} the base coordinates $u,v$ can no longer vanish simultaneously implying that their associated divisors do not intersect
\be\label{eq:BBIntersectionRelation}
\lb S_u-E\rb \cdot \lb S_v-E\rb=0\,.
\ee
The exceptional divisor $E$ is a curve of self-intersection $-1$ while $S_u,S_v$ intersect exactly once in $B_2$ and thus we derive
\be
S_u\cdot E= S_v\cdot E=0\,,
\ee
which together with the previous result fixes all intersection of the kind \eqref{eq:Intersections}. 

For horizontal divisors in the 3-fold $Z_3$ we have the relations 
\be
\sigma \cdot  D_{u_i}=\sigma \cdot  D_{v_i}=\sigma \cdot  D_{\delta_i}=0\,,
\ee
as the centers of the blowups introducing the exceptional divisors $D_{u_i},D_{v_i},D_{\delta_i}$ are located in the $w=1$ patch of $Z_3$. Finally note that the intersection of any three divisors $D_i$ in $Z_3$ can be lifted to an intersection in the ambient space $Z_4$ by
\be
D_1\cdot  |_{Z_3\:}D_2\cdot  |_{Z_3\:}D_3= D_1\cdot  |_{Z_4\:}D_2\cdot  |_{Z_4\:}D_3\cdot |_{Z_4\:} [Z_3]\,.
\ee

Each blowup introduces a projective relation prohibiting the involved coordinates from vanishing simultaneously. Consequently their associated divisors do not intersect. One therefore obtains a quadratic or cubic intersection relation for every blowup. For example the projective relation $[x:y:w]$ gives rise to the identity
\be\ba
0&=\sigma \cdot \lb 2\sigma + 2c_1\rb \cdot \lb 3\sigma+3c_1 \rb\,,
\ea\ee
while the base blowup gives rise to \eqref{eq:BBIntersectionRelation}. These relations can be used to evaluate mixed intersections involving distinct horizontal divisors \cite{Lawrie:2012gg}.

A subset of the exceptional divisors $D_{u_i},D_{v_i}$ are Cartan divisors $D_{\alpha_i}^{E_n}$ labelled by the root $\alpha_i$ they correspond to within the affine $E_n$ root system. On the 3-fold $Z_3$ these are fibered as
\be\label{eq:Fibration}
\P^1_{\alpha_i}\hookrightarrow D_{\alpha_i}^{E_n}\xrightarrow[]{\pi} W_{\nu}\,,
\ee
where $\nu=u,v$ depending on which coordinate is associated with the $E_n$ singularity. We equivalently write $D_{\alpha_i}^{E_n}=D_{\alpha_i}^{(\nu)}$ if the $E_n$ singularitiy is fibered above $\nu=0$. Here we introduced $W_\nu= \lbbb \nu =0\rbbb  \subset Z_3  $. The position of the fibers $\P^1_{\alpha_i}$ within the exceptional locus $\pi^{-1}(W_\nu)$ are determined by the pull back properties of the projection $\pi:Z_3 \rightarrow \widetilde B_2$, which are derived from the replacement relations \eqref{eq:BaseBlowUp} and \eqref{eq:BlowUps2} and condensed in the transformations 
\be\ba\label{eq:ReplacementRelations}
u &\rightarrow u \, \epsilon^{\xi_1^{(u)}}\, \prod_{j=2}^{r}\delta_j^{\xi_j^{(u)} }\prod_{i}u_i^{m_i^{(u)}}\,, \\
 v &\rightarrow v \, \epsilon^{\xi_1^{(v)}}\, \prod_{j=2}^{r}\delta_j^{\xi_j^{(v)}} \prod_{i}v_i^{m_i^{(v)}}\,, \\  \epsilon &\rightarrow \epsilon \prod_{j=2}^{r}\delta_j^{\zeta_j}\,.
\ea\ee
Here $m_i^{(u,v)}$ are the Dynkin labels of the respective Lie algebra $E_n,E_m$ and we take the relations \eqref{eq:ReplacementRelations} as a definition for the multiplicity integers $\xi_j^{(u,v)}$. This implies that the pull backs of the vanishing loci $W_\nu$ and the exceptional base divisor $E$ are given by
\be\ba\label{eq:Multiplicities}
\pi^{-1}\lb W_u \rb &= \sum_{k=1}^r \xi_k^{(u)}S_k  +\sum_i m_i^{(u)}D_{u_i}\,, \\
\pi^{-1}\lb W_v \rb &= \sum_{k=1}^r \xi_k^{(v)}S_k  +\sum_i m_i^{(v)}D_{v_i}\,, \\
\pi^{-1}(E)&= \sum_{k=2}^r \zeta_kS_k
\ea\ee
where the divisors $\lbbb \nu_i=0\rbbb = D_{\nu_i}$ and surfaces  $S_k=\lbbb \delta_k=0 \rbbb$ for $k=2,\dots,r$ and $S_1=\lbbb \epsilon =0\rbbb $ with $\nu=u,v$ have been introduced. 

The fiber $\P^1_{\alpha_i}$ of a Cartan divisor $D_{\alpha_i}^{E_n}, D_{\alpha_i}^{E_m}$ is called a flavor curve if it is fully contained in the reducible surface $S=\cup_{k=1}^r S_k$. Consider a divisor $ D_{\alpha_i}^{E_n}$ associated to the complex base coordinate $u$ and therefore giving $W_u$ when projected to the base. Then the intersection $\pi^{-1}(W_v)\cdot D_{\alpha_i}^{E_n}=\P^1_{\alpha_i}$ yields the full fiber component of the horizontal divisor $D_{\alpha_i}^{E_n} $ as $\pi^{-1}(W_v)$ is vertical and $W_u,W_v$ intersect transversely in the base. For Cartan divisors $D_{\alpha_i}^{(u)}$ associated to simply laced algebras one therefore has
\be
\pi^{-1}(W_v)\cdot D_{\alpha_i}^{(u)} \cdot D_{\alpha_i}^{(u)} =-2\,,
\ee
where we have indicated the coordinates $(u,v)$ the algebra $E_n$ is associated with by superscripts rather than the algebra. Intersecting the expansion of $\pi^{-1}(W_v)$ given in \eqref{eq:Multiplicities} with $D_{\alpha_i}^{(u)} \cdot D_{\alpha_i}^{(u)}$ we find two negative contributions. The intersection with $\sum_i m_i^{(v)}D_{v_i}$ vanishes precisely when the fiber of $D_{\alpha_i}^{(u)}$ is a flavor curve as then it is not contained any of the $D_{v_i}$. Thus $D_{\alpha_i}^{(u)}$ contributes a flavor curve presicely when
\be\label{eq:Int}
\sum_{k=1}^r \xi_k^{(v)}S_k \cdot D_{\alpha_i}^{(u)} \cdot D_{\alpha_i}^{(u)}=-2\,.
\ee
Identical arguments apply with $u,v$ interchanged. 

\subsection{Reduced Intersection Matrices}
\label{sec:RedIntMat}

The reduced intersection matrices encode how the flavor curves $\P^1_{\alpha_i}$ \eqref{eq:Fibration} of the Cartan divisors $D_{\alpha_i}^{E_n}$ are contained within the surface components $S_k$.  It is motivated by \eqref{eq:Int} and defined by the triple intersections  
\be\label{eq:ReducedInt}
I_{ik}^{E_n}= D_{\alpha_i}^{E_n} \cdot  D_{\alpha_i}^{E_n}\cdot  S_k\,,
\ee
within $Z_3$. The criterium \eqref{eq:Int} for a Cartan divisor to contribute a flavor curve now becomes
\be\ba
\tn{Flavor Curve }\P^{1}_{\alpha_i}\subset D^{E_n}_{\alpha_i}\,:\qquad \sum_{k=1}^r \xi^{(u,v)}_iI_{ik}^{E_n}=-2\,,\\
\ea\ee 
where intersection are weighted by $\xi^{(u,v)}$ if the divisors $D^{E_n}_{\alpha_i}$ project to $W_v,W_u$ respectively. Directly from \eqref{eq:ReplacementRelations} using the blowups \eqref{eq:blowups} and \eqref{eq:ResolutionSequenceMarginal} we compute the integers $\xi^{(u,v)}$ to
\be\label{eq:WeightsExplicit}
\ba
\xi^{(u)}_{(E_6,E_7)}&=(1,1,2,1,3,1,1,2,1,2,1)\,, \\  \xi^{(v)}_{(E_6,E_7)}&=(1,1,1,2,1,3,1,2,1,1,2)\,,\\
\xi^{(u)}_{(E_6,E_8)}&=(1, 1, 2, 1, 3, 1, 4, 5, 1, 1, 2, 2, 3, 2, 2, 4, 1, 1, 3, 3, 2)\,,\\
\xi^{(v)}_{(E_6,E_8)}&=(1, 1, 1, 2, 1, 3, 1, 1, 1, 1, 2, 2, 3, 1, 1, 2, 2, 2, 1, 2, 3)\,,\\
\xi^{(u)}_{(E_7,E_8)}&=(1, 1, 2, 1, 3, 1, 4, 1, 1, 1, 2, 1, 2, 3, 2, 2, 4, 1, 1, 2, 3, 1, 3,
  2)\,,\\
\xi^{(v)}_{(E_7,E_8)}&=(1, 1, 1, 2, 1, 3, 1, 4, 5, 1, 2, 1, 2, 3, 1, 1, 2, 2, 2, 4, 1, 3, 2,
  3)\,.\\
\ea
\ee
Taking these multiplicities into account we find the three tables of triple intersections
\renewcommand{\arraystretch}{1.25}
\setlength{\arraycolsep}{4pt}
\be\ba\label{tab:Int}
&\begin{array}{c|ccccccccccccccc}
	D_{\alpha_i}^2 S_k &  v & v_8 & v_6 & v_7 & v_5 & v_3 & v_2 & u & u_{6} & u_{11} & u_{10} & u_{9} & u_{7} & u_{4} & u_{8} \\\hline
	\epsilon & -2 & -2 & -2 & -1 & 0 & -1 & 0 & -2 & -2 & -2 & -2 & -1 & 0 & 0 & -1  \\
	\delta _5 & 0 & 0 & 0 & 0 & 0 & 0 & 0 & 0 & 0 & 0 & 0 & -1 & -2 & -2 & -1 \\
	\delta _6 & 0 & 0 & 0 & -1 & -2 & -1 & -2 & 0 & 0 & 0 & 0 & 0 & 0 & 0 & 0 \\
	\sum \xi_kS_k & -2 & -2 & -2 & -2 & -2 & -2 & -2 & -2 & -2 & -2 & -2 & -2 & -2 & -2 & -2  \\
\end{array}\\
&\begin{array}{c|cccccccccccccccc}
	D_{\alpha_i}^2 S_k &  v & v_7 & v_8 & v_6 & v_5 & v_3 & v_2 & u & u_{10} & u_9 & u_{12} & u_{13} & u_{15} & u_{14} & u_7 & u_{8} \\\hline
	\epsilon & -2 & -2 & -2 & -1 & 0 & -1 & 0 & -2 & -2 & -2 & -2 & -2 & -2 & -1 & 0 & -1 \\
	\delta _6& 0 & 0 & 0 & -1 & -2 & -1 & -2 & 0 & 0 & 0 & 0 & 0 & 0 & 0 & 0 & 0 \\
	\delta _8& 0 & 0 & 0 & 0 & 0 & 0 & 0 & 0 & 0 & 0 & 0 & 0 & 0 & -1 & -2 & -1  \\
	\sum \xi_kS_k & -2 & -2 & -2 & -2 & -2 & -2 & -2 & -2 & -2 & -2 & -2 & -2 & -2 & -2 & -2 & -2 \\
\end{array} \\
&\begin{array}{c|ccccccccccccccccc}
	D_{\alpha_i}^2 S_k &  u & u_6 & u_{11} & u_{10} & u_9 & u_7 & u_4 & u_8 & v & v_{8} & v_{7} & v_{11} & v_{13} & v_{14} & v_{15} & v_{9} & v_{10} \\\hline
	\epsilon & -2 & -2 & -2 & -2 & -1 & 0 & 0 & -1 & -2 & -2 & -2 & -2 & -2 & -1 & 0 & 0 & 0  \\
	\delta _7 & 0 & 0 & 0 & 0 & -1 & -2 & -2 & -1 & 0 & 0 & 0 & 0 & 0 & 0 & 0 & 0 & 0  \\
	\delta _9 & 0 & 0 & 0 & 0 & 0 & 0 & 0 & 0 & 0 & 0 & 0 & 0 & 0 & -1 & -2 & -2 & -2  \\
	\sum \xi_kS_k & -2 & -2 & -2 & -2 & -2 & -2 & -2 & -2 & -2 & -2 & -2 & -2 & -2 & -2 & -2 & -2 & -2 \\
	\end{array}
\ea\ee

\noindent for $(E_6,E_7),(E_6,E_8), (E_7,E_8)$ respectively. We only depict the non-vanishing rows. The final row lists the summed columns weighted with multiplicities. The weights $\xi_k^{(u,v)}$ for all rows depicted is $1$ as seen in \eqref{eq:WeightsExplicit}. The final row verifies that all Cartan divisors contribute a flavor curve to the geometry making it marginal.

\subsection{Fiber Diagrams and Surface Geometries}
\label{sec:Fiberdiagrams}

\begin{figure}
  \centering
  \includegraphics[width=13cm]{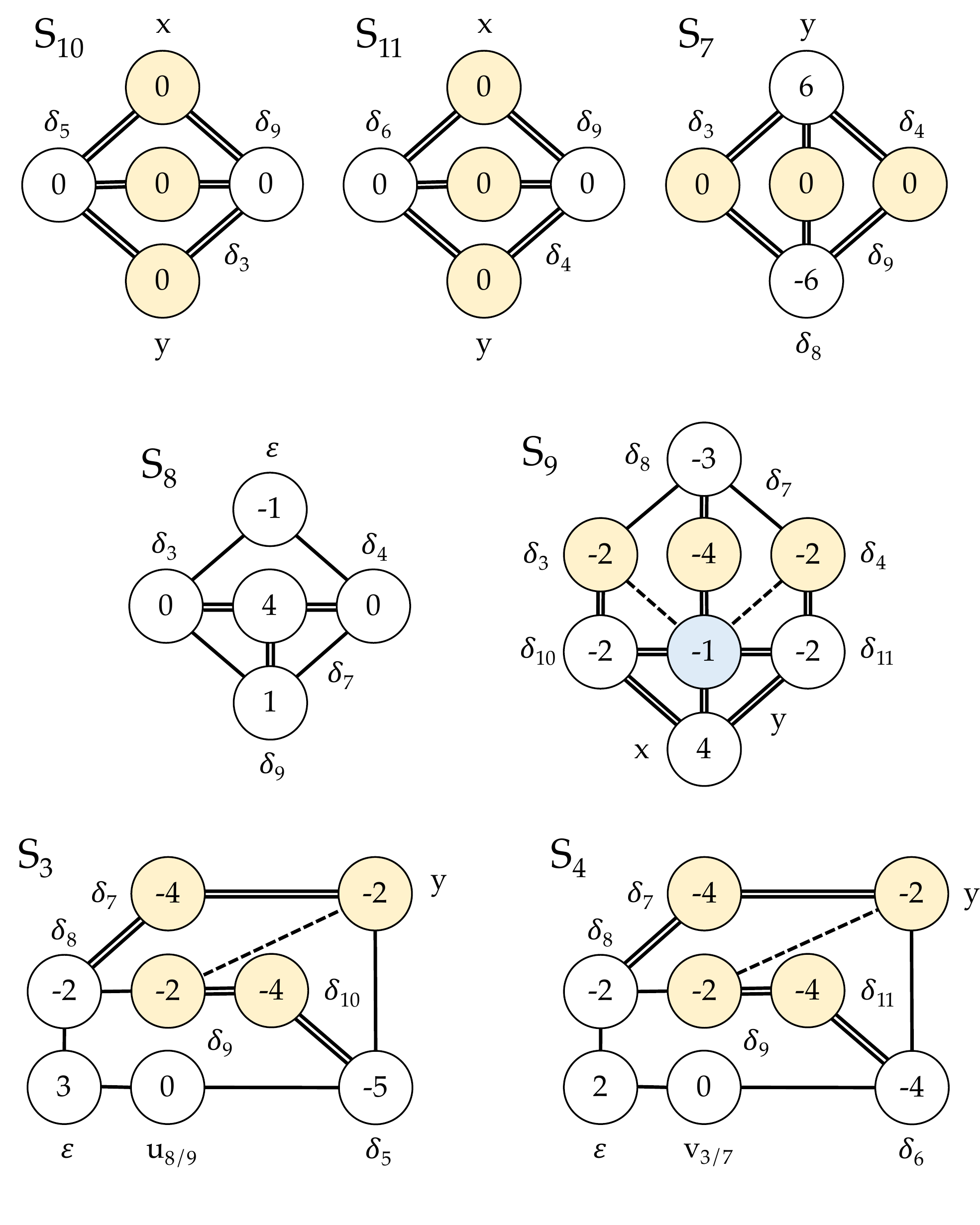}
    \caption{The figure shows the intersection matrix \eqref{eq:IntMatrix} for the rank 10 $(E_6,E_7)$ marginal geometry computed from the resolution \eqref{eq:blowups}. The genus of a yellow/blue curves is formally $g=-1,-2$ respectively, while uncolored curves are of vanishing genus. The former are reducible while the latter are irreducible. Dashed lines denote negative intersection between curves indicating common components.}\label{fig:E6E7Raw1}
\end{figure} 

\begin{figure}
  \centering
  \includegraphics[width=13cm]{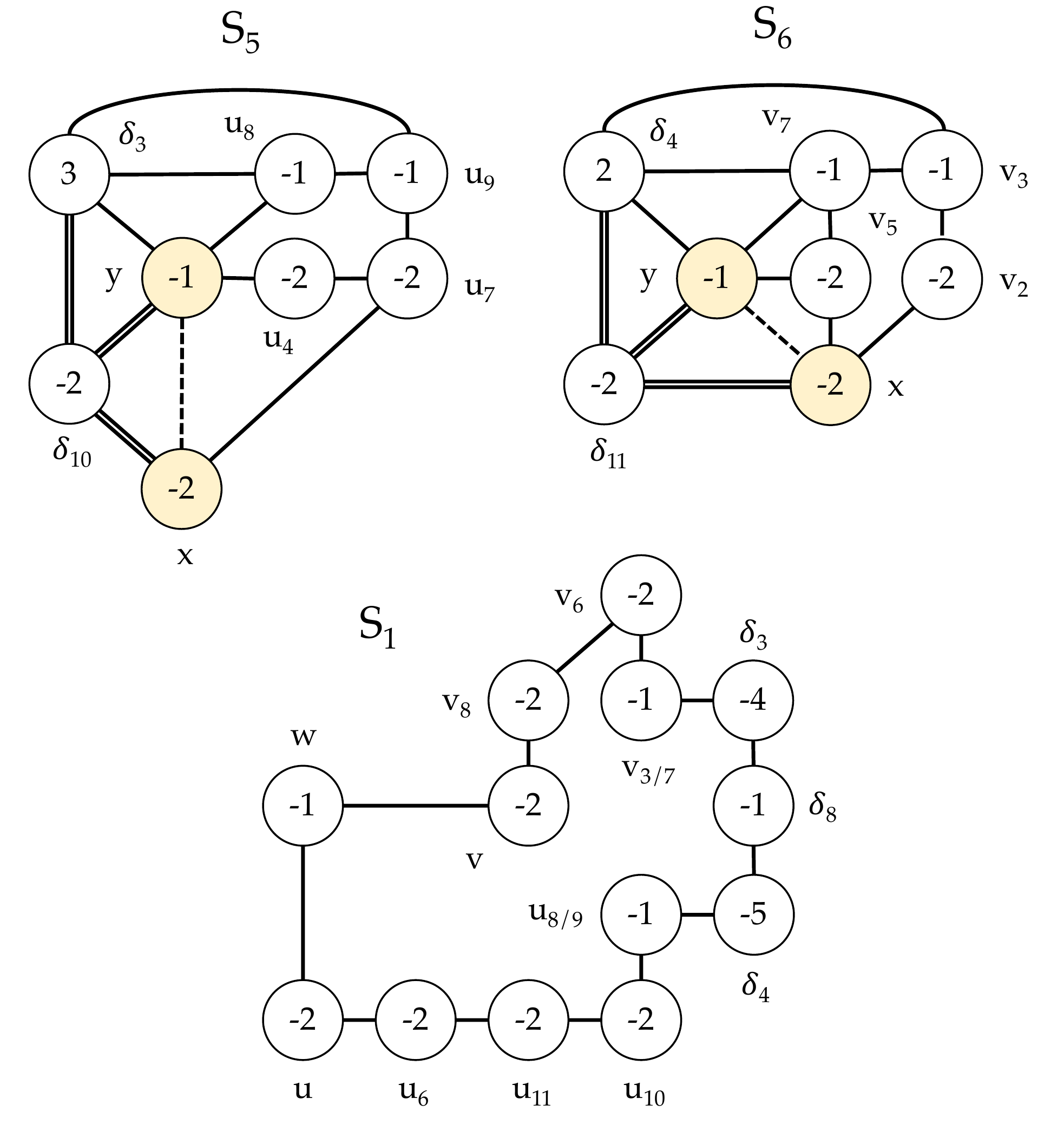}
    \caption{Continuation of figure \ref{fig:E6E7Raw1}.}\label{fig:E6E7Raw2}
\end{figure} 

In the singular limit $Z_3\rightarrow X_3$ the SCFT originates from the collapse of the reducible surface $S=\cup_{k} S_k$ in the non-flat fibration. M5 branes wrapping irreducible components of $S$ give rise to tensionless magnetically charged strings and M2 branes wrapping curves in $S$ generate a tower of electrically charged states. Both enhance the spectrum in the singular limit. The nature of these enhancements depends crucially on the geometry of $S=\cup_k S_k$ which we study here in detail for the geometries of $(E_n,E_m)$ conformal matter given in \eqref{sec:EnEmGeometries}. Much of the SCFT data can be captured in so called combined fiber diagrams (CFDs) which effectively depict the Mori-cone of $S$ \cite{Apruzzi:2019vpe,Apruzzi:2019opn, Apruzzi:2019enx} together with curves of vanishing self-intersection. They subsume the resolution dependent fiber diagrams associated to each surface component $S_k$ into a single graph.

Fiber diagrams are pictorial representations of the full intersection matrix
\be\label{eq:IntMatrix}
I_{pqk}= D_p \cdot  D_q\cdot S_k\,,
\ee 
within $Z_3$ where $D_{p,q}$ run over all Cartan divisors $D_{\alpha_i}^{E_n}, D_{\alpha_j}^{E_m}$ together with the divisors $D_x,D_y,D_w,D_{\delta_k}$ associated to the sections $x,y,w,\delta_k$. Each surface $S_k$ has its own fiber diagram and depicts the intersection matrix \eqref{eq:IntMatrix} for fixed $k$. Nodes represent divisors and are labelled by the associated sections. The self-intersection of each curve $S_k\cdot D_p$ is recorded in the center of the representing node. The genus of a curve $C\subset S_k$ is fixed from the intersection matrix by the relation
\be
K_{S_k}\cdot|_{S_k\,} C+C\cdot|_{S_k\,} C=2g(C)-2=S_k\cdot S_k \cdot C+C\cdot C \cdot S_k\,,
\ee
with the double intersections taken in the surface $S_k$ and the triple intersections taken in the Calabi-Yau 3-fold. Here $K_{S_k}$ is the canonical bundle of the surface.

We now explicitly discuss the rank 10 $(E_6,E_7)$ marginal geometry, giving the results for the other cases of $(E_n,E_m)$ without derivation as the analysis extends unaltered to these cases. The fiber diagrams for a rank 10 $(E_6,E_7)$ marginal geometry are shown in figures \ref{fig:E6E7Raw1} and \ref{fig:E6E7Raw2}. The genus of the uncolored curves vanishes, the light yellow curves have a formal genus of $g=-1$ and the single light blue curve has a formal genus of $g=-2$  indicating that these are reducible. Solid links denote positive intersection between two curves, dashed links denote negative ones indicating that connected surfaces share common irreducible components. Finally the degree of the surface $S_k$, given by the self-intersection of the canonical divisor, is given by the triple intersection number
\be
S_k^3=(-1,4,4,5,4,8,8,4,8,8)\,,
\ee
for $k=1,3,4,\dots,11$. The leading $(-1)$ follows as the surface $S_1$ in figure \ref{fig:E6E7Raw2} is the degree 9 surface $\P^2$ blown up 10 times. 

While fiber diagrams are readily computable they only depict the intersection relations between the curves $D_p\cdot S_k$ which for higher rank geometries are generally reducible and given by a linear combinations of curves generating the geometry of the generalized del Pezzo surfaces $S_k$. We now make the connection to the classification of generalized del Pezzo surfaces as given in \cite{Derenthal2014SingularDP} manifest. We begin by flopping $v_{3/7},v_6,v_8$ and $u_{8/9},u_{10},$ $u_{11},u_6$ from $S_1$ over $S_3,S_4$ to $S_5,S_6$ respectively to collect all the curves contributing to the $E_6,E_7$ Dynkin diagrams in the surfaces $S_5,S_6$. These flops amount to changing the order of the blowups \eqref{eq:blowups} and a transition of phases in the gauge theory description. Blowing down a (-1) curve connecting to curves of self-intersection $m,n$ gives the transition
\be\label{eq:BlowDown}
m-(-1)-n \quad \rightarrow \quad (m+1)-(n+1)\,,
\ee
while blowing up (-1) curve has the opposite effect
\be\label{eq:BlowUp}
m-n \quad \rightarrow \quad (m-1)-(-1)-(n-1)\,.
\ee
Here flopping a curve from $S_k$ into $S_l$ involves a blow down in $S_k$ and a blowup in $S_l$. Further blowups and blow downs decrease and increase the degrees of the surfaces by 1 such that the degrees of $S_1,S_5,S_6$ are now $6,1,1$ respectively. The reduced intersection matrix for the flopped geometry reads
\renewcommand{\arraystretch}{1.25}
\be\ba\label{tab:Int2}
&\begin{array}{c|ccccccccccccccc}
	D_{\alpha_i}^2 S_k &  v & v_8 & v_6 & v_7 & v_5 & v_3 & v_2 & u & u_{6} & u_{11} & u_{10} & u_{9} & u_{7} & u_{4} & u_{8} \\\hline
	\epsilon & -1 & 0 & 0 & 0 & 0 & 0 & 0 & -1 & 0 & 0 & 0 & 0 & 0 & 0 & 0  \\
	\delta _5 & 0 & 0 & 0 & 0 & 0 & 0 & 0 & -1 & -2 & -2 & -2 & -2 & -2 & -2 & -2 \\
	\delta _6 & -1 & -2 & -2 & -2 & -2 & -2 & -2 & 0 & 0 & 0 & 0 & 0 & 0 & 0 & 0 \\
	\sum \xi_kS_k & -2 & -2 & -2 & -2 & -2 & -2 & -2 & -2 & -2 & -2 & -2 & -2 & -2 & -2 & -2  \\
\end{array}
\ea\ee

\begin{figure}
  \centering
  \includegraphics[width=11cm]{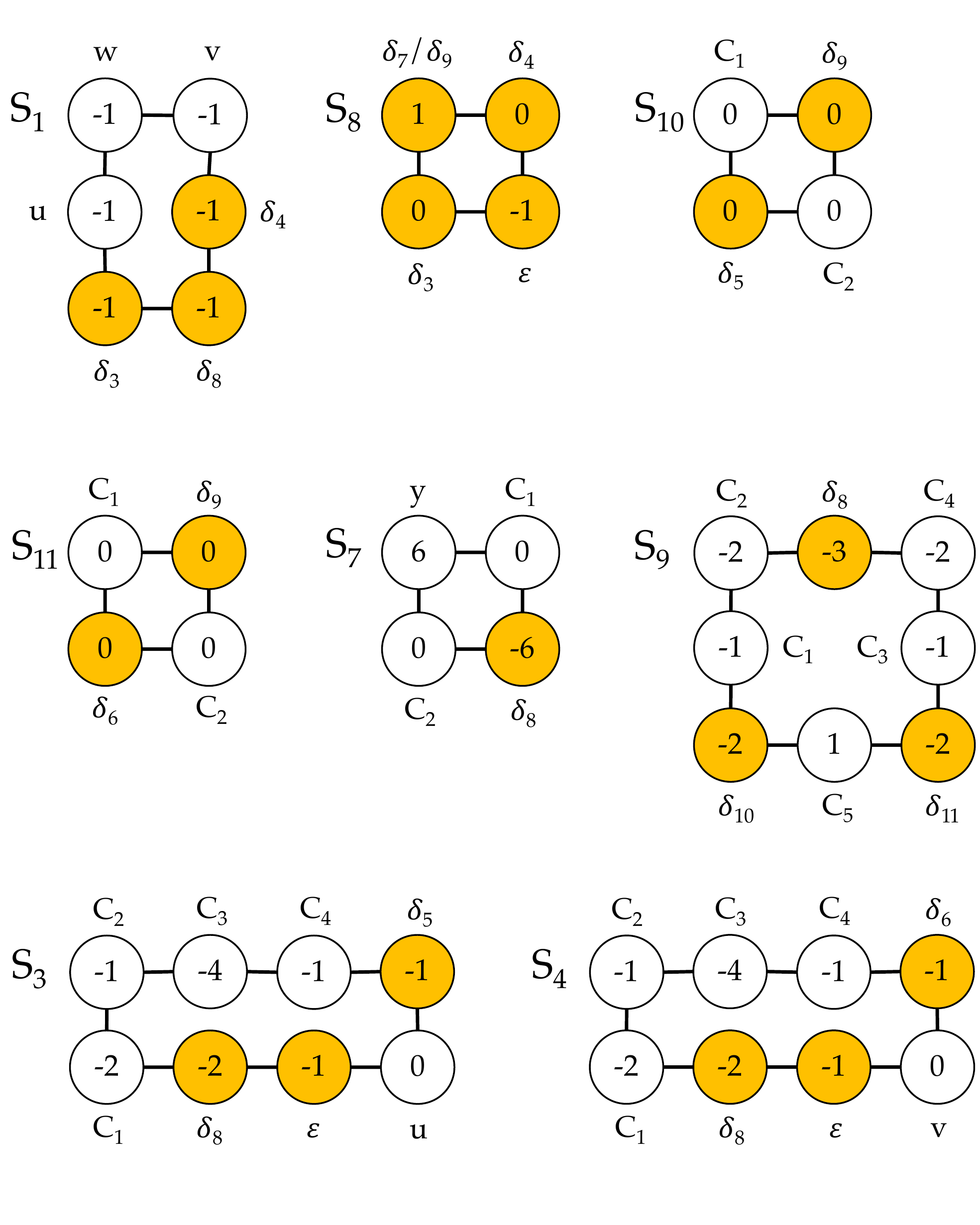}
    \caption{The picture shows an $(E_6,E_7)$ marginal geometry related to the blowup sequence in \eqref{eq:blowups}. Each diagram depicts the full set of generators for the Cox ring of the surface $S_k$. The curves $C_i$ in individual surfaces are distinct and enumerate excess generators not directly associated to section of the Calabi-Yau geometry. Their relation to the divisors of the Calabi-Yau restricted to $S_k$ is listed in \eqref{eq:DetailedRelations}. Flavor curves of self-intersection $(-2)$ are colored green, manifest gluing curves are colored yellow and the remaining curves are colored white. Homologous curves are listed by `$/\,$'.}\label{fig:FDE6E71}
\end{figure} 

\begin{figure}
  \centering
  \includegraphics[width=11cm]{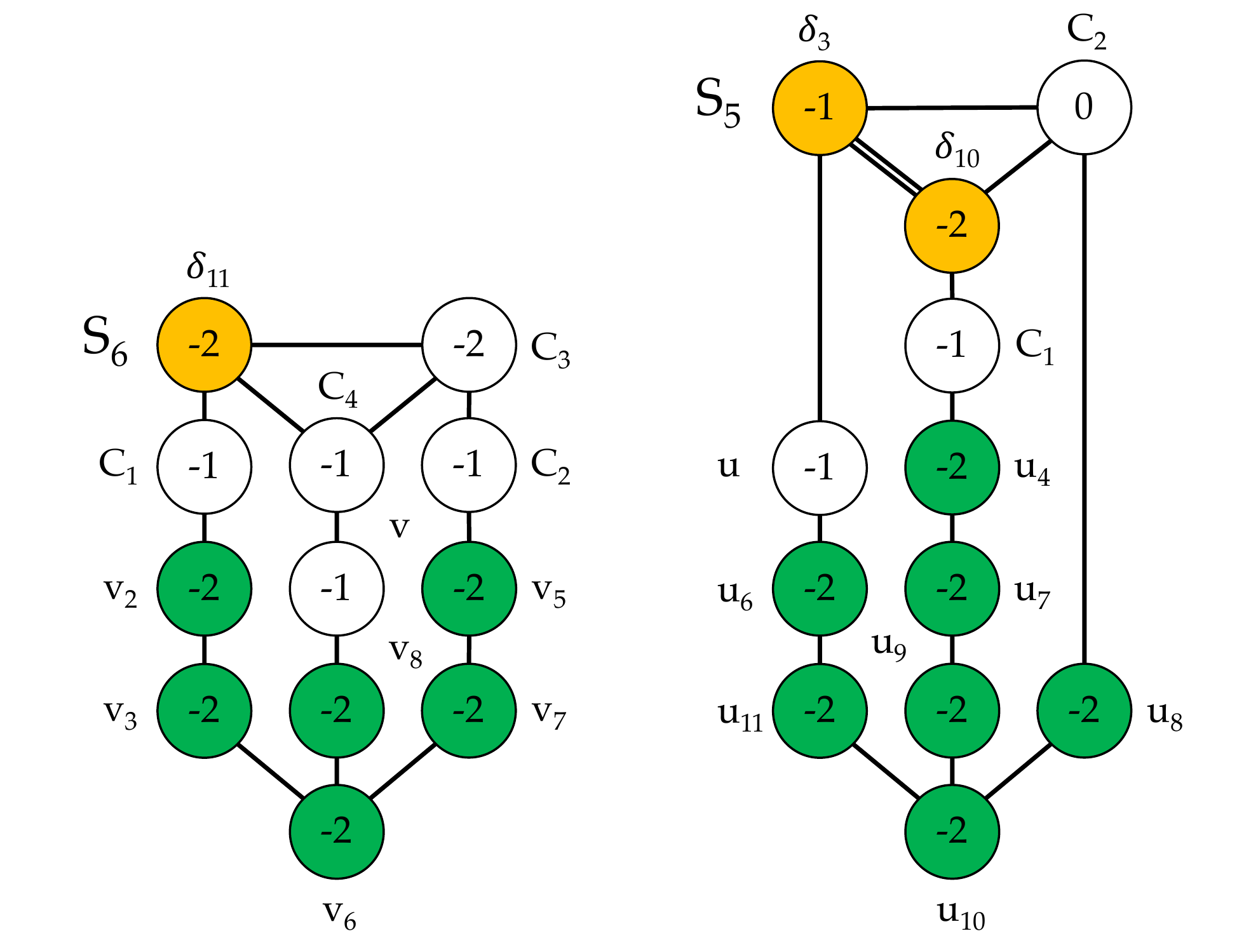}
    \caption{$(E_6,E_7)$ marginal geometry. Figure \ref{fig:FDE6E71} continued.}\label{fig:FDE6E72}
\end{figure} 

\noindent Next we decompose the reducible curves into irreducible components by explicitly studying the hypersurface equation restricted to the locus of the reducible curve.  These are then matched to the geometries of generalized del Pezzo surfaces as classified by degree in \cite{Derenthal2014SingularDP}. We determine the geometries, shown in figures \ref{fig:FDE6E71} and  \ref{fig:FDE6E72}, of the individual surfaces underlying the fiber diagrams to be
\be\ba
S_1^6\,:&\qquad \tn{Bl}_3\P^2\quad &&\tn{(toric)}\\
S_3^4\,:&\qquad \tn{Bl}_5\P^2\quad &&\tn{(toric)}\\
S_4^4\,:&\qquad \tn{Bl}_5\P^2\quad &&\tn{(toric)}\\
S_5^1\,:&\qquad \tn{Type}\,(D_6+A_1)_1\quad &&\tn{(non-toric)} \\
S_6^1\,:&\qquad \tn{Type}\,(A_5+A_2)_1\quad &&\tn{(non-toric)}\\
S_7^8\,:&\qquad \mathbb{F}_6&&\tn{(toric)}\\
S_8^8\,:&\qquad \mathbb{F}_1&&\tn{(toric)}\\
S_9^4\,:&\qquad \tn{Bl}_5\P^2&&\tn{(toric)}\\
S_{10}^8\,:&\qquad \P^1\times \P^1&&\tn{(toric)}\\
S_{11}^8\,:&\qquad \P^1\times \P^1 &&\tn{(toric)}\\
\ea\ee
where we have denoted the degree of the surfaces as given in figures  by a superscript and $\mathbb{F}_1,\mathbb{F}_6$ are Hirzebruch surfaces. The subscripts on $(D_6+A_1)_1,(A_5+A_2)_1$ denote the number of blow-ups performed on the geometry. The intersection of the fiber diagrams shown in figures \ref{fig:E6E7Raw1} and \ref{fig:E6E7Raw2} are reproduced by the  identifications\medskip
\be\label{eq:DetailedRelations}
\ba
S_3\,: &\quad \delta_7=2C_1+2C_2+C_3\,,\quad \delta_9=C_1+2C_2\,,\quad \delta_{10}=C_3+2C_4\,, \quad y=C_2+C_4\,,\\
S_4\,: &\quad \delta_7=2C_1+2C_2+C_3\,, \quad \delta_9=C_1+2C_2\,,\quad \delta_{11}=C_3+2C_4\,, \quad y=C_2+C_4\,, \\
S_5\,:&\quad x=2C_1+u_4\,,\quad  y=C_1+C_2\,,\\
S_6\,:&\quad x=C_1+C_2+C_3\,, \quad y=C_1+C_2+C_3+v+v_8+v_6+v_3+v_2\,, \quad \delta_4=C_3+C_4\\
S_7\,:&\quad \delta_3=2C_1\,, \quad \delta_4=2C_2\,,\quad \delta_9=C_1+C_2\,,\\
S_9\,:&\quad \delta_3=2C_1+C_2\,, \quad \delta_4=2C_3+C_4\,,\quad \delta_7=C_2+C_4\,,\quad x=2C_5\,,\quad y=C_1+C_3+C_5\,,\\
S_{10}\,:&\quad \delta_3=2C_1\,, \quad x=2C_2\,, \quad y=C_1+C_2\,, \\
S_{11}\,:&\quad \delta_4=2C_1\,, \quad x=2C_2\,, \quad y=C_1+C_2\,, \\
\ea
\ee
in the geometries shown in figures \eqref{fig:FDE6E71} and  \eqref{fig:FDE6E72}. Here we have only listed the identification of the reducible green curves in figures \ref{fig:E6E7Raw1} and \ref{fig:E6E7Raw2}. Curves $C_i\subset S_k$ are a priori unrelated and enumerate excess generators of the Cox ring of $S_k$ not directly associated with a section of the resolved Calabi-Yau geometry.

\subsection{Combined Fiber diagrams (CFDs)}

\begin{figure}
  \centering
  \includegraphics[width=11cm]{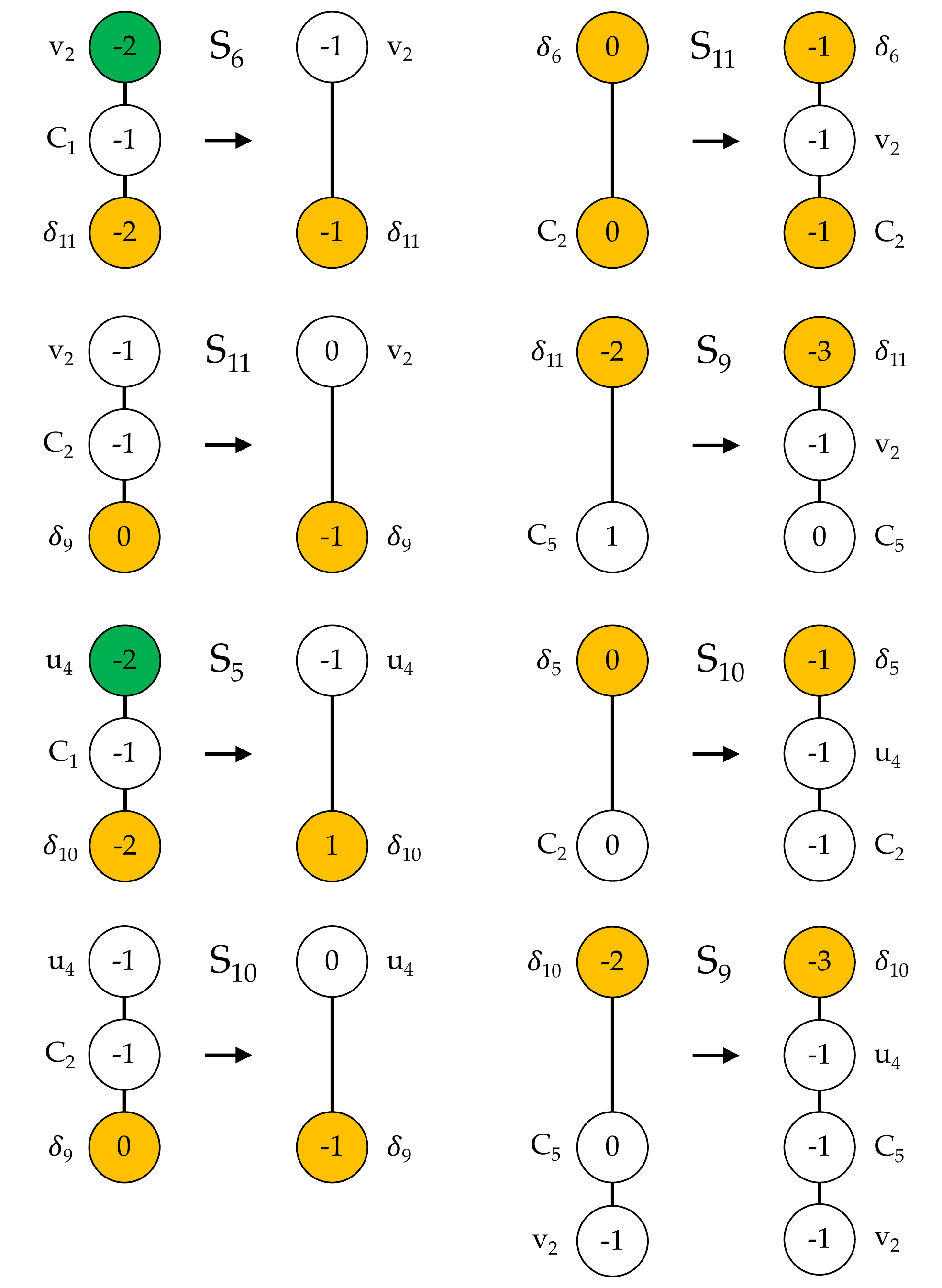}
    \caption{In each line the figure shows a pair of blow downs and blowups used to partially flop $v_2,u_4$ from the surfaces $S_5,S_6$ over $S_{11},S_{10}$ to $S_9$. This series of flops makes the $(-1)$ curve of the CFD which connects to $u_4,v_2$ manifest.}\label{fig:Flops}
\end{figure} 

A combined fiber diagram (CFD) is generated from a collection of fiber geometries associated to the surface components $S_k$ by jointly representing the Mori-cone generators of the surfaces $S_k$, i.e. the $(-2)$ curves and rational $(-1)$ curves with normal bundle $\CO(-1)\oplus\CO(-1)$. Gluing curves $\Sigma_{kl}=S_k\cap S_l$ are excluded. In addition curves of vanishing self-intersection are shown. The CFDs are independent of the flop transitions moving curves between the $S_k$ and describe all Coulomb branch phases of a gauge theory equally \cite{Apruzzi:2019enx}. 

In general the $(-1)$ curves will be a linear combination of the curves depicted in figure \ref{fig:FDE6E71} and \ref{fig:FDE6E72}, however manipulating the geometry using the flop transitions we can make these manifest. For the $(E_6,E_7)$ geometry these flops are depicted in figure \ref{fig:Flops} with the resulting $(-1)$ curve labelled by $C_5$ emerging in the surface component $S_9$ as shown in the bottom right of figure \ref{fig:Flops}. A second $(-1)$ curve is already manifest within the surface component $S_1$ shown in figure \ref{fig:FDE6E71} and labelled by $w$. The $(-2)$ curves are the flavor curves colored green in figures \ref{fig:FDE6E71} and \ref{fig:FDE6E72}.

\begin{figure}
  \centering
  \includegraphics[width=6cm]{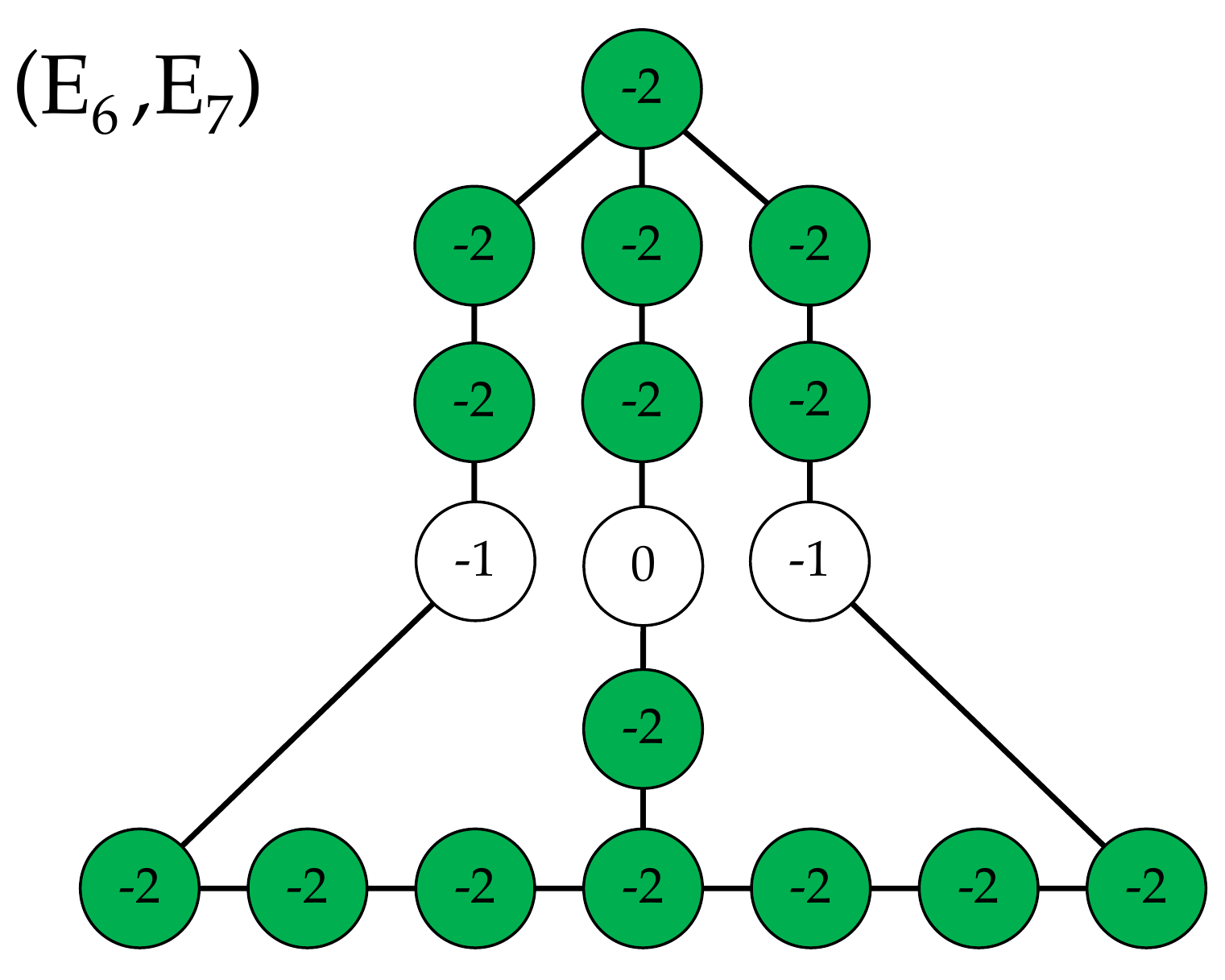}
    \caption{Marginal Combined fiber diagram (CFD) for $(E_6,E_7)$ conformal matter. The genus of all depicted curves vanishes. }\label{fig:CFDE6E7}
\end{figure}

\begin{figure}
  \centering
  \includegraphics[width=7cm]{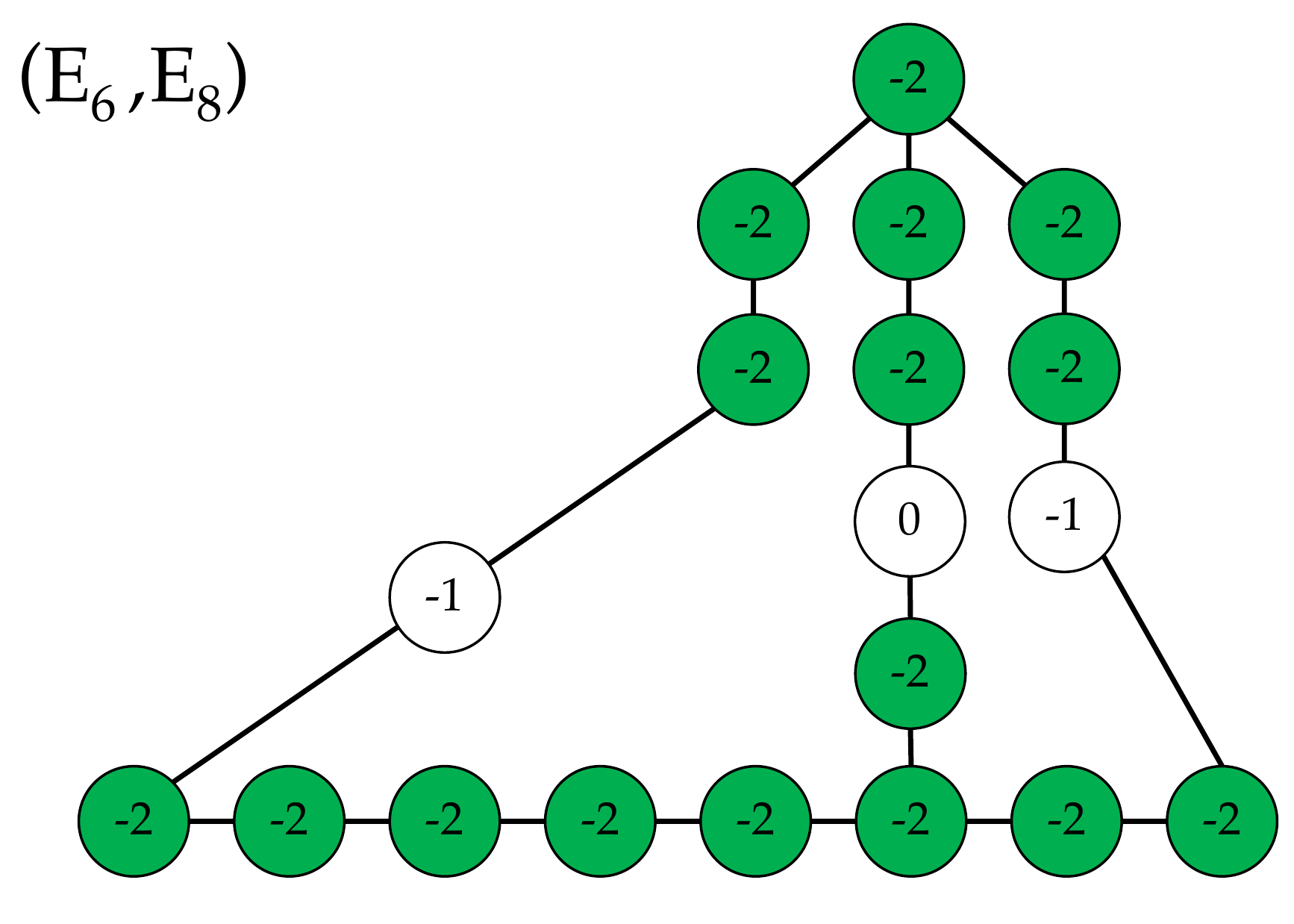}
    \caption{Marginal Combined fiber diagram (CFD) for $(E_6,E_8)$ conformal matter.  The genus of all depicted curves vanishes. }\label{fig:CFDE6E8}
\end{figure}

\begin{figure}
  \centering
  \includegraphics[width=8cm]{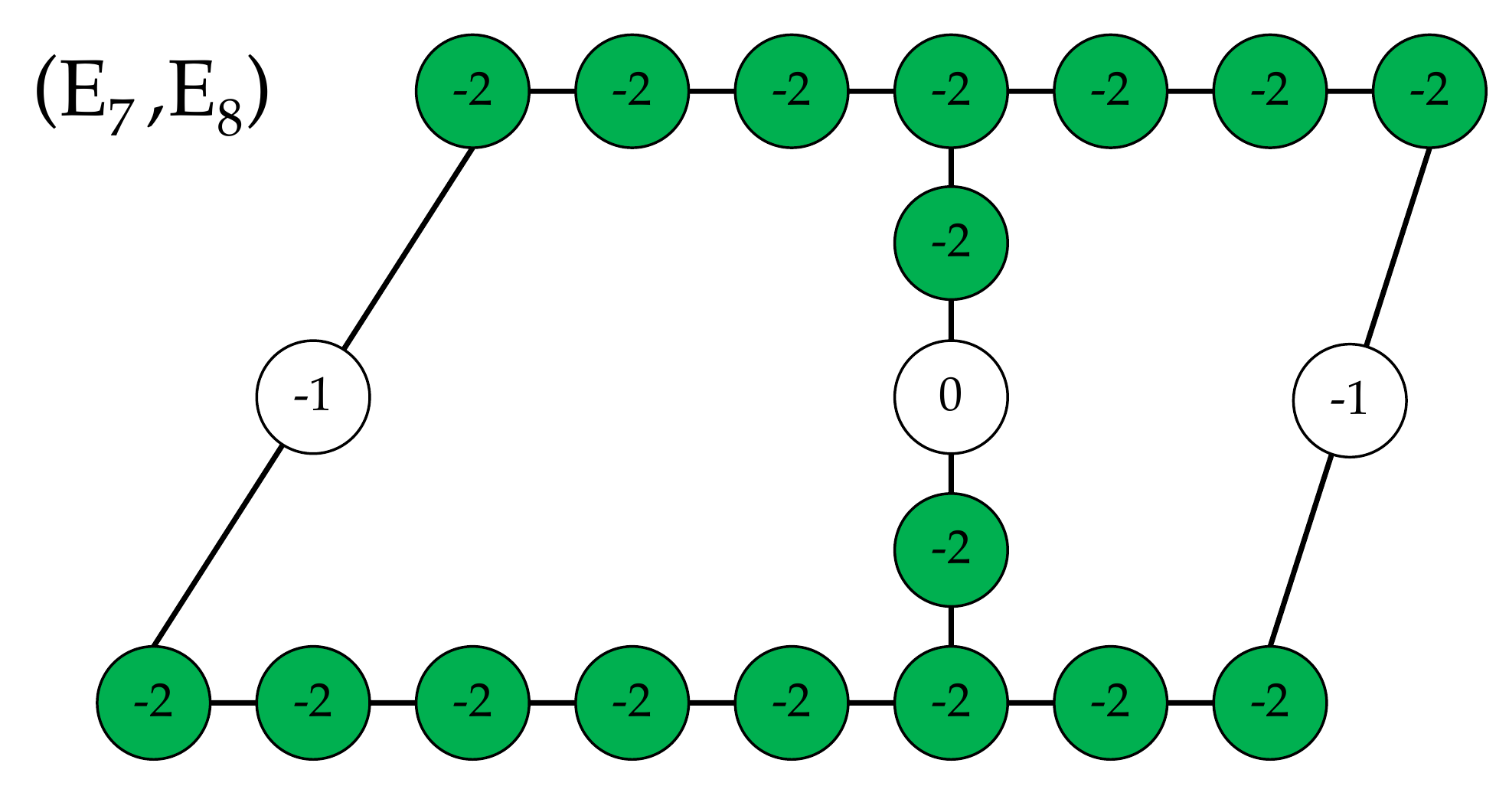}
    \caption{Marginal Combined fiber diagram (CFD) for $(E_7,E_8)$ conformal matter.  The genus of all depicted curves vanishes.}\label{fig:CFDE7E8}
\end{figure} 

Alternatively we can compute the reduced intersection matrices
\be
I_{zk}^{(E_n,E_m)}= D_{z} \cdot   D_{z} \cdot S_k\,,
\ee
with $z=x,y,w$ and sum over all surface components $S_k$. In all cases $(n,m)=(6,7),(6,8),(7,8)$ we find
\be\ba
\sum_{k}I_{xk}^{(E_n,E_m)}&=0\,,\\
\sum_{k}I_{yk}^{(E_n,E_m)}&=-1\,,\\
\sum_{k}I_{wk}^{(E_n,E_m)}&=-1\,,\\
\ea\ee indicating the presence of two $(-1)$ curves within $S=\bigcup_{i=1}^rS_k$. From the fibers given in figures \ref{fig:FDE6E71} and \ref{fig:FDE6E72} one extracts the flavor curves these connect to and the presence of an additional curve of vanishing self-intersection. 

We depict the three resulting CFDs associated to $(E_6,E_7),(E_6,E_8),(E_7,E_8)$ conformal matter in figures \ref{fig:CFDE6E7}, \ref{fig:CFDE6E8}, \ref{fig:CFDE7E8}. All depicted curves are curves of vanishing genus and the normal bundle of the (-1) curves is given by $\CO(-1)\oplus\CO(-1)$.

\section{Descendants and Weakly Coupled Quivers} 
\label{sec:DescendantsAndQuivers}

The combined fiber diagrams (CFDs) derived in section \ref{sec:MarginalGeometries} distill key features of the resolutions \eqref{eq:ResolutionSequenceMarginal} and \eqref{eq:blowups}. Taking the CFDs in figures \ref{fig:CFDE6E7}, \ref{fig:CFDE6E8} and \ref{fig:CFDE7E8} as starting points we turn to discuss descendant 5d SCFTs and weakly coupled quiver descriptions of $(E_n,E_m)$ conformal matter.

\subsection{Descendant SCFTs}

\begin{figure}
  \centering
  \includegraphics[width=6cm]{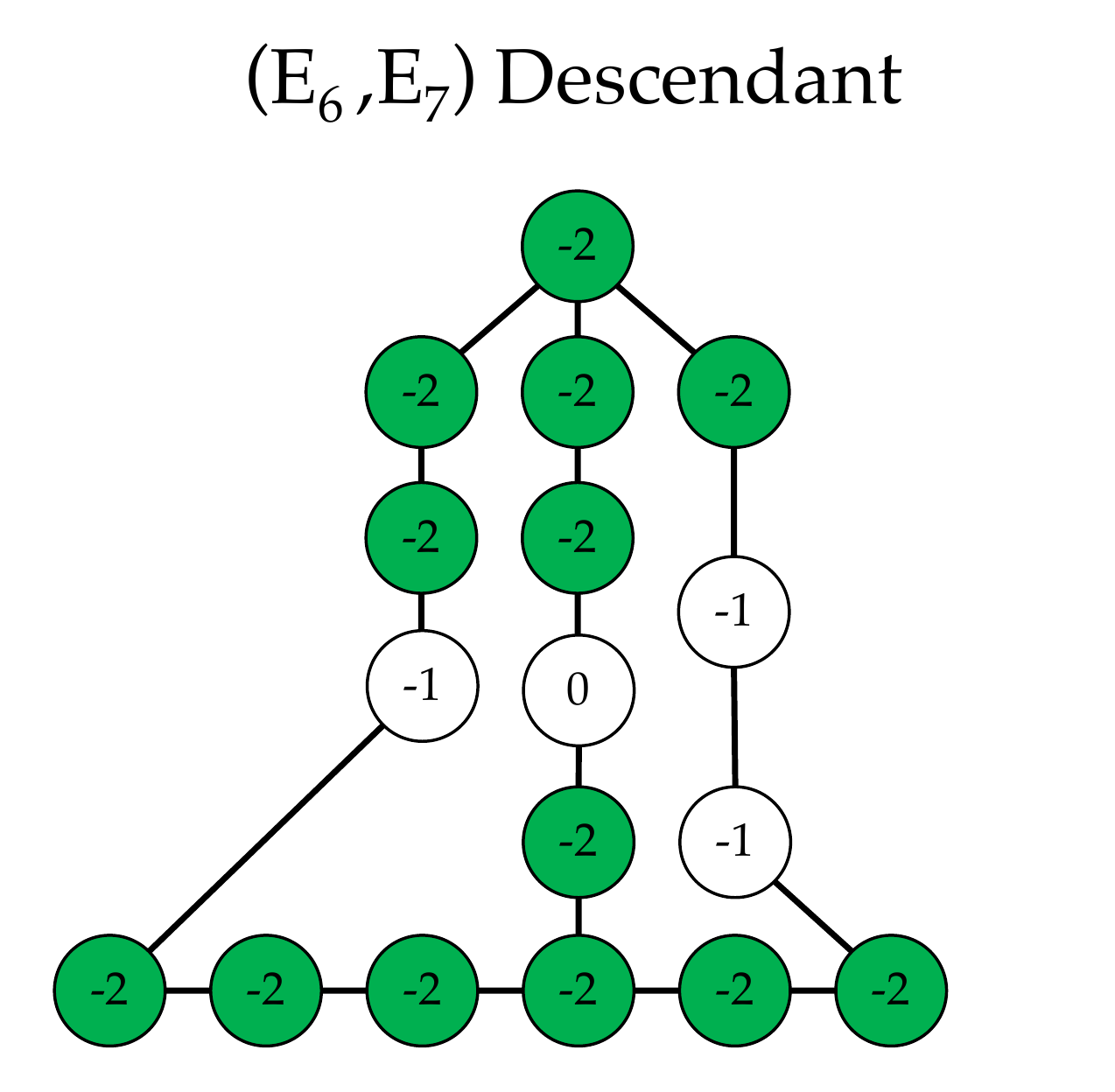}
    \caption{The CFD of the first descendant of the $(E_6,E_7)$ marginal CFD derived from a single application of the CFD transition rules \eqref{eq:Adjacencyupdate}, \eqref{eq:LabelUpdate} to the right most (-1) curve of the marginal CFD in figure \ref{fig:CFDE6E7}. }\label{fig:DescendentExample}
\end{figure} 

The structure of the collapsing surface $S=\cup_k S_k$ determines the SCFT data in the singular limit and can be manipulated in various ways. In section \ref{sec:Fiberdiagrams} we flopped $(-1)$ curves between surface components $S_k$ to access different gauge theory phases of the weakly coupled description of the marginal gauge theory. Alternatively $(-1)$ curves can be flopped out of $S$ resulting in a complex surface $S'$ which is not phase equivalent to the one it originates from. The SCFT generated by M-theory when collapsing the surface $S'$ is referred to as a descendant theory or just descendant of the SCFT associated to the surface $S$. Its flavor symmetry, BPS spectrum and weakly coupled descriptions are distinct from the marginal theory. The series of flop transitions giving all possible descendant theories of a marginal theory can be unified into unique manipulations on its associated CFD referred to as CFD transitions which were laid out in \cite{Apruzzi:2019vpe, Apruzzi:2019opn} and which we now reproduce. 

Denote the nodes of a CFD by $C_i$ and label these with their self-intersection and genus $(n_i,g_i)$. The intersection matrix is denoted by $m_{ij}=C_i\cdot C_j$. Given this data a CFD transition generates a new CFD given by the labels $(n_i',g_i')$ and the intersection matrix $m_{ij}'$ with the two rules: 
\begin{enumerate}
\item Remove a curve $C_i$ of self-intersection (-1) and vanishing genus from the CFD, delete the corresponding row and column of the intersection matrix $m_{ij}$ and update the reduced matrix $m_{jk}'$ according to
\be\label{eq:Adjacencyupdate}
m'_{jk}=m_{jk}+m_{ij}m_{ik}
\ee
with $i\neq j, k$. If $C_i$ intersects multiple curves apply the rule \eqref{eq:Adjacencyupdate} pairwise.
\item Update the labels of the remnant curves $C_j$ according to
\be\label{eq:LabelUpdate}
n'_j=n_j+m_{ij}^2\,, \qquad g_j'=g_j+\frac{m_{ij}^2-m_{ij}}{2}\,.
\ee
\end{enumerate}
We give an example of the CFD transition generating the first descendant of the marginal $(E_6,E_7)$ geometry in figure \ref{fig:DescendentExample}.

The full tree of descendants is obtained by applying the CFD transition rules \eqref{eq:Adjacencyupdate}, \eqref{eq:LabelUpdate} until there are no more (-1) curves remaining in the CFD. Enumerating the list of descendants for $(E_n,E_m)$ conformal matter we thus find
\be
(E_6,E_7)\,:~ 90\,, \qquad (E_6,E_8)\,:~ 196\,, \qquad (E_7,E_8)\,:~ 225\,.
\ee
descendant CFDs and 5d SCFTs.

\subsection{Constraints on Weakly Coupled Quiver Descriptions}
\label{sec:Constraints}

\begin{figure}
  \centering
  \includegraphics[width=7cm]{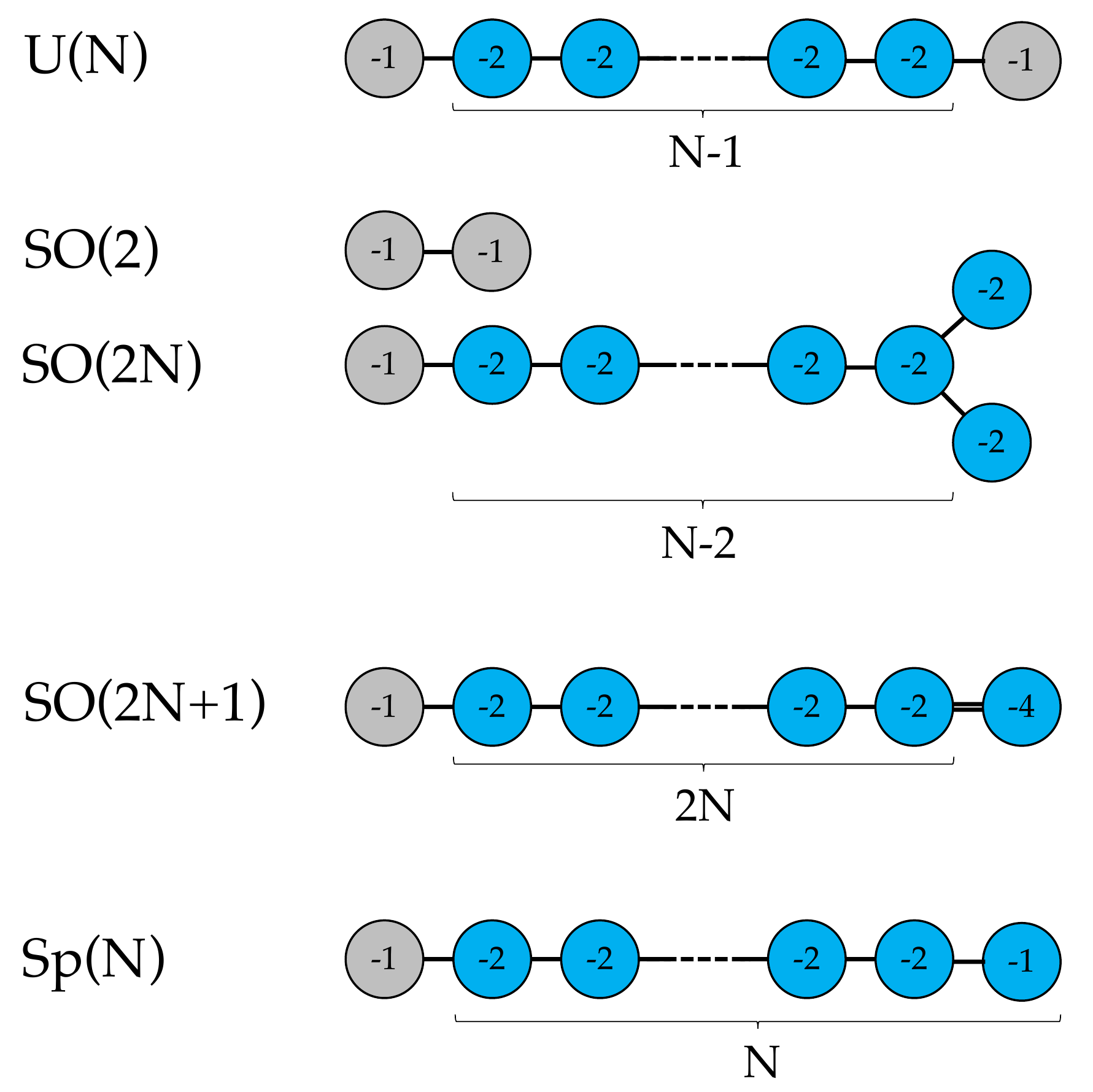}
    \caption{List of box graph CFDs derived from the extended Coulomb branch of weakly coupled 5d quiver gauge theories. The integer $N$ enumerate the number of blue curves of the box graph CFD. These necessarily embed into the CFD of any 5d SCFT to which the quiver theory completes. Only $U(N)$ and $SU(N)$ box graph CFDs embed into the marginal CFDs of $(E_n,E_m)$ conformal matter. The embeddings are shown in figure \ref{fig:BGCFDEmbedded}. This list is a partial recreation of a table found in \cite{Apruzzi:2019enx}.}\label{fig:BGCFDs}
\end{figure} 

Given a marginal 5d theory its possible weakly coupled quiver gauge theory descriptions are heavily constrained:
\begin{enumerate}
\item Box Graph CFDs: The flavor symmetry of a quiver gauge theory is a subgroup of the flavor symmetry of the SCFT it completes to. Further, descendants of quiver gauge theories are weakly coupled descriptions of the descendants of the associated SCFT and the structure of the extended Coulomb branch must embed within the marginal geometry. As a consequence box graph CFDs derived from the extended Coulomb branch must form subgraphs of the CFD \cite{Apruzzi:2019enx}. Conversely, the possible subgraphs of the marginal CFD correspond to partial quivers embedded within any consistent quiver gauge theory completing to the SCFT. The list of box graph CFDs is given in figure \ref{fig:BGCFDs}. When multiple subgraphs are embedded they must not intersect. This ensures that the descendant structure of the quiver gauge theory is reproduced within that of the SCFTs.

\item Gauge and Flavor Rank: The gauge rank $r_G$ of a quiver gauge theory is given by the sum of the ranks of the gauge nodes and must coincide with the rank for the SCFT, i.e. the number of irreducible surface components of $S=\cup_kS_k$ as they were counted in \eqref{eq:Ranks}. The flavor rank $r_F$ is the rank of the total global symmetries. For quivers the total global symmetry receives a factor of the topological $U(1)_I$ abelian symmetry for every gauge node and a factor $U(1)_B$ for a single full hypermultiplet in the bifundamental of two gauge groups. Finally the classical flavor symmetries contribute. The flavor rank coincides with the rank of the enhanced flavor symmetry of the SCFT and for $(E_n,E_m)$ conformal matter is simply $r_F=n+m+1$.

\item Number of Hypermultiplets: The number of hypermultiplets connecting to any single gauge node is constrained by positivity conditions on the Coulomb branch metric and monopole string tensions \cite{Jefferson:2017ahm}. We list the implied restrictions on the matter content relevant for weakly coupled quiver descriptions of $(E_n,E_m)$ conformal matter in table \ref{tab:NumberConstraints}.
\end{enumerate}

\noindent We now repeatedly apply these rules to determine quiver candidates for $(E_n,E_m)$ conformal matter theories.

\subsection{Quiver Descriptions of Maximal and Submaximal Depth}
\label{sec:QuiversE6E7}

We now derive quiver descriptions consistent with the conditions above for $(E_n,E_m)$ conformal matter theories. We discuss each theory in turn.

\begin{figure}
  \centering
  \includegraphics[width=15.5cm]{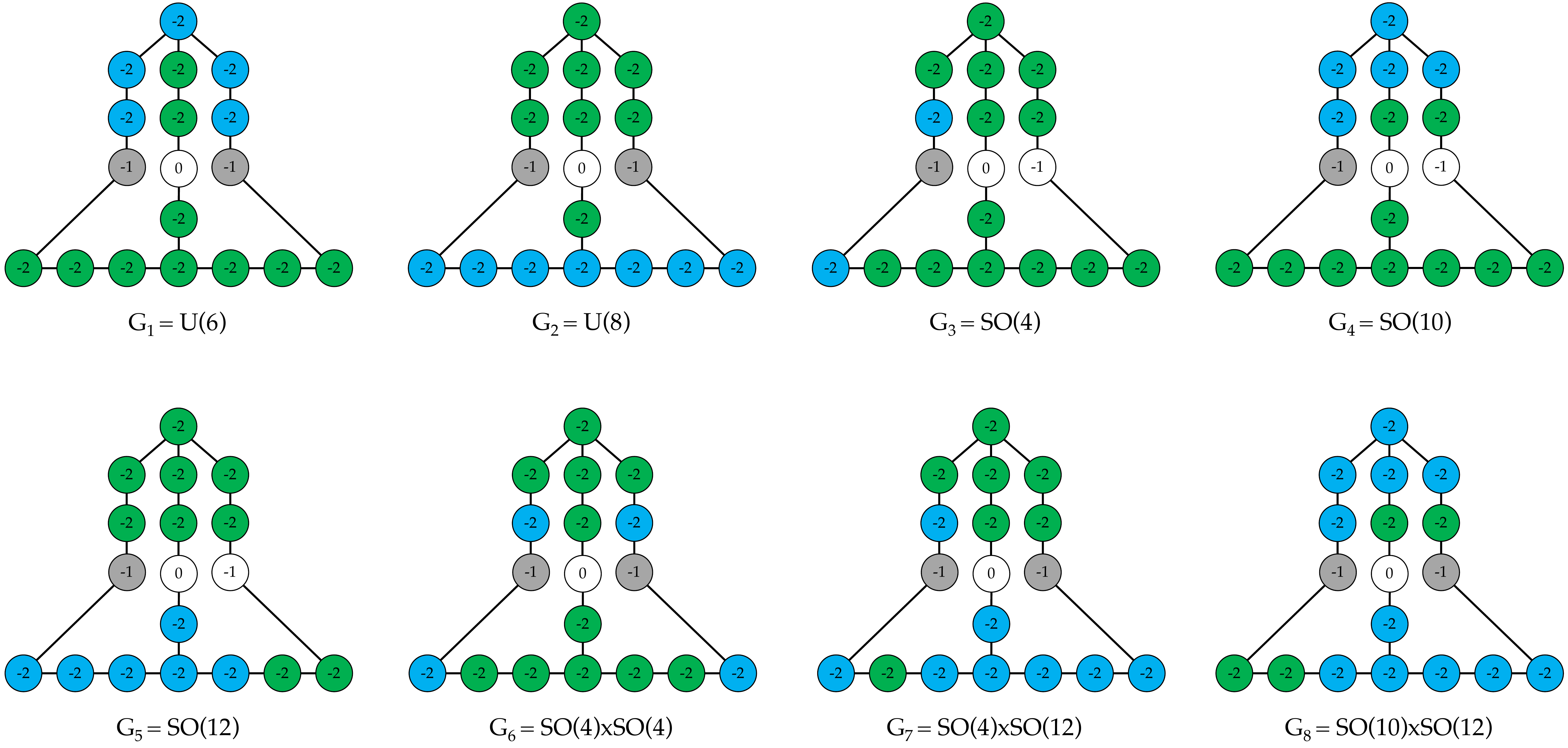}
    \caption{List of possible embeddings of box graph CFDs into the marginal CFD of $(E_6,E_7)$ conformal matter. Four embeddings related by a symmetry to the embeddings 3,4,5,7,8 are omitted. Each graph is labelled with the flavor symmetry manifest in the corresponding subquiver. When multiple box graph CFDs are embedded no two curves of distinct embeddings intersect. }\label{fig:BGCFDEmbedded}
\end{figure}

From the box graphs in figure \ref{fig:BGCFDs} only those of type $U(N)$ and $SO(2N)$ embed into the marginal CFD of figures \ref{fig:CFDE6E7}, \ref{fig:CFDE6E8}, \ref{fig:CFDE7E8}. The possible eight embeddings into the marginal $(E_6,E_7)$ CFD are shown in figure \ref{fig:BGCFDEmbedded} together with the flavor symmetry they make manifest. Each of these embeddings gives rise to a subquiver which realizes this flavor symmetry as rotations on its hypermultiplets, the pairs are listed in table \ref{tab:BauSteine}. 

Next we connect the consistent subquivers by introducing additional gauge nodes and bifundamental hypermultiplets. The resulting quiver must have the gauge and flavor rank
\be
(E_6,E_7)\,:\quad \lb r_G,r_F\rb=\lb 10 ,14 \rb\,,
\ee
be anomaly free and respect the consistency constraints on the number of attached hypermultiplets at each gauge node as listed in table \ref{tab:NumberConstraints}. There are many such quivers and we restrict the analysis to those of maximal depth, i.e. those with the most descendants or equivalently with the highest number of matter multiplets. 

\renewcommand{\arraystretch}{1.55}
\begin{table}
\begin{center}
 \begin{tabular}{||c | c  ||}
 \hline
 Subgroup of $G_F$ & Subquivers for $(E_6,E_7)$ \\ [0.5ex] 
 \hline\hline
 $U(6)$ & $6{\bf F}-SU(n)$  \\ 
 \hline
 $U(8)$ & $8{\bf F}-SU(n)$ \\
 \hline
 $SO(4)$ & $2{\bf F}-Sp(n)$ \\
 \hline
 $SO(10)$ & $5{\bf F}-Sp(n)$  \\
 \hline
  $SO(12)$ & $6{\bf F}-Sp(n)$  \\
 \hline
 $SO(4)\times SO(4)$ & $ 2{\bf F}-Sp(n_1)\oplus  2{\bf F}-Sp(n_2)$ \\
  \hline
 $SO(4)\times SO(12)$ & $ 2{\bf F}-Sp(n_1)\oplus  6{\bf F}-Sp(n_2)$ \\
  \hline
 $SO(10)\times SO(12)$ & $ 5{\bf F}-Sp(n_1)\oplus  6{\bf F}-Sp(n_2)$ \\
  [1ex] 
 \hline
\end{tabular}
\end{center}
\caption{We list the flavor subgroups and their corresponding subquiver for $(E_6,E_7)$ conformal matter derived from the possible embeddings of the box graph CFDs in figure \ref{fig:BGCFDEmbedded} into the marginal $(E_6,E_7)$ CFD.}
\label{tab:BauSteine}
\end{table}

\renewcommand{\arraystretch}{1.55}
\begin{table}
\begin{center}
 \begin{tabular}{||c | c  | c ||}
 \hline 
 Quiver Gauge Group & Representations & Upper Bounds  \\ [0.5ex] 
 \hline\hline
 $SU(N)$ & $({\bf Sym},{\bf AS},{\bf F};k)$ & $\begin{array}{c}  (1,1,1;0),(1,0,N-2;0),(1,0,0;N/2), \\ (0,2,8;0),(0,2,7;3/2),(0,1,N+6;0), \\ (0,1,8;N/2),(0,0,2N+4;0), \end{array}$  \\ 
 \hline
 $Sp(N)$ & $({\bf AS},{\bf F})$ & $(1,8),(0,2N+6)$\\
  \hline
 $SO(N)$ & $({\bf V})$ & $(N-2)$\\
   \hline
 $F_4$ & $({\bf 26})$ & $(3)$\\
 \hline
 \hline
 $SU(3)$ & $({\bf F};k)$ & $(6;4),(3;13/2),(0;9)$ \\
 \hline
 $SU(4)$ & $({\bf AS},{\bf F};k)$ & $\begin{array}{c}  (4,0;4),(3,4;2),(3,0;5),(2,0;6), \\ (1,0;7),(0,8;3),(0,0;8) \end{array}$  \\
 \hline
  $SU(5)$ & $({\bf AS},{\bf F};k)$ & $(3,3;0),(3,1;3),(3,2;3/2),(0,5;11/2) $ \\
  \hline
   $SU(6)$ & $({\bf AS},{\bf F};k)$ & $(3,0;3)$ \\
 \hline
 $Sp(2)$ & $({\bf AS},{\bf F})$ & $(3,0),(2,4)$  \\
 \hline
  $Sp(3)$ & $({\bf AS},{\bf F})$ & $(2,0)$ \\
 \hline
\end{tabular}
\end{center}
\caption{The configuration of hypermultiplets connecting to a gauge node of a  5d quiver gauge theory are constrained if it is to complete to an SCFT. The table summarizes the constraints for $SU(n),Sp(n)$ gauge nodes including their low rank outliers. The second column abbreviates the notation used in the third. Here ${\bf Sym},{\bf AS},{\bf F},{\bf V},k$ denote the symmetric, antisymmetric, fundamental and vector representations as well as the Chern-Simons level respectively. For $Sp(n),SO(n),F_4$ the final column is an upper bound on the possible number hypermultiplets while for $SU(n)$ the interpretation is more subtle, we refer to \cite{Jefferson:2017ahm}. The above table is a partial recreation of tables found in \cite{Apruzzi:2019enx}.}
\label{tab:NumberConstraints}
\end{table}

These quivers are of the structure
\be\ba\label{eq:possibleQuivers2}
\tn{Maximal Depth Quivers }(E_6,E_7)\,:\qquad 5{\bf F}-Sp(n_1)-\prod G-Sp(n_2)-6{\bf F}\,,
\ea\ee
with $n_1+n_2+n_{\Pi G}=10$ where $\prod G$ abbreviates the internal structure of the 5d quiver and $n_{\Pi G}$ the sum of the ranks of the gauge nodes it features. The links in \eqref{eq:possibleQuivers2} connecting to the fundamental matter are full hyper-multiplets.

The global symmetry rank must be 14 whereby the interior is either empty and the $Sp(n)$ gauge nodes are connected by a full bifundamental hypermultiplet or it consists of a single gauge node $G$ linked to the $Sp(n)$ gauge nodes by half-hypermultiplets. Consequently such a node $G$ must have a real fundamental representation. This in turn poses the additional constraint that this representation should be even dimensional as the theory is otherwise anomalous, due to an $Sp(n)$ gauge node connecting to a total odd number of half-hypermultiplets. 

We consider the case of an empty interior first and connect the two symplectic gauge groups by a single full bifundamental hypermultiplet. The global symmetry group of this quiver is
\be
G=SO(10)\times SO(12)\times U(1)_I^2 \times SU(2)_B\,,\qquad r_F=14\,,
\ee 
where the $SU(2)_B$ rotates the two bifundamental half-hyper multiplets.
The global symmetry rank is as required. The remaining constraints then read
\be\ba
Sp(n_1)\,:&&\qquad\quad 2n_2+5&\leq 2n_1+6\,, \\
Sp(n_2)\,:&&\qquad\quad 2n_1+6&\leq 2n_2+6\,, \\
r_G\,:&&\qquad\quad n_1+n_2&=10\,,  
\ea \ee
where the first two inequalities are derived from the constraint on the number of fundamental hypermultiplets as listed in table \ref{tab:NumberConstraints} for the gauge nodes $Sp(n_1),Sp(n_2)$ respectively. The only admissible quiver of this type is thus
\be\label{eq:CandidateQuiver}
5{\bf F}-Sp(5)-Sp(5)-6{\bf F}\,.
\ee
The last consistency condition we apply is that the classical global symmetries of the descendants of quiver \eqref{eq:CandidateQuiver} must include into the flavor group of the corresponding CFD descendants. The first descendant of the marginal $(E_6,E_7)$ CFD is shown in figure \ref{fig:DescendentExample} and displays the global symmetry
\be
G=E_6\times E_7\times U(1)_I\,,
\ee
while the two corresponding descent quivers of the candidate \eqref{eq:CandidateQuiver}, obtained by decoupling one hypermultiplet on either side of the quiver, display the global symmetry groups
\be\ba\label{eq:GlobalSymmetries}
G_1&=SO(10)\times SO(10)\times U(1)\times U(1)\times SU(2)\,, \\
G_2&=SO(8)\times SO(12)\times U(1)\times U(1)\times SU(2)\,.
\ea\ee
We have that $SO(10)\times U(1)\subset G_1$ and $SO(10)\times SU(2)\subset G_1$ as well as that $SO(8)\times U(1)\subset G_2$ and $SO(12)\times SU(2)\subset G_2$ include into $E_6,E_7$ respectively. The quiver \eqref{eq:CandidateQuiver} is thus a good weakly coupled quiver candidate for $(E_6,E_7)$ conformal matter.

We move on to study the case where the interior of the maximal depth quivers \eqref{eq:possibleQuivers} consists of a single gauge node $G$ which we require to have a real, even dimensional fundamental representation. This leaves the two choices $G=F_4, SO(2n)$ for the added gauge node. We consider the case $G=F_4$ first. The constraints in table \ref{tab:NumberConstraints} now yield the bounds
\be\ba\label{eq:ConstraintsF4}
Sp(n_1)\,:&&\qquad\quad 5+\frac{26}{2}&\leq 2n_1+6\,,\\
Sp(n_2)\,:&&\qquad\quad 6+\frac{26}{2}&\leq 2n_2+6\,,\\
F_4\,:&&\qquad\quad \frac{2n_1}{2}+\frac{2n_2}{2}&\leq 3\,,\\
\ea\ee
with the gauge rank constraint $n_1+n_2=6$. The gauge rank constraint together with the last constraint in \eqref{eq:ConstraintsF4} are not solvable and $G=F_4$ is excluded. When $G=SO(2n)$ the constraints take the form
\be\ba\label{eq:ConstraintsSO}
Sp(n_1)\,:&&\qquad\quad 5+n&\leq 2n_1+6\,,\\
Sp(n_2)\,:&&\qquad\quad 6+n&\leq 2n_2+6\,,\\
SO(2n)\,:&&\quad n_1+n_2&\leq 2n-2\,,\\
\ea\ee
with the gauge rank constraint $n_1+n_2+n=10$. This system has four solutions given by $(n_1,n_2,n)=(2,3,5),(2,4,4),(3,3,4),(4,2,4)$. For each of these quivers the global symmetry of all of its descendants is a subgroup of the global symmetry of the associated SCFT descendants derived from the CFD transitions. 

It follows with the same reasoning as above that the last class of maximal depth quivers, which are of the structure 
\newcommand{\bminus}{\mathbin{\rotatebox[origin=c]{90}{$-$}}}  
\renewcommand{\arraystretch}{0.9}
\setlength{\arraycolsep}{1.75pt}
\be\ba\label{eq:Quiver22}
\begin{array}{ccccccc}
& & G & & & & \\
& & \bminus & & & &  	\\
	5{\bf F} & - & Sp(n_1) &  - & Sp(n_2) & - & 6{\bf F} \\
\end{array} \qquad\quad \textnormal{or}\qquad\quad  \begin{array}{ccccccc}
& & & & G & & \\
& & & & \bminus & &  	\\
	5{\bf F} & - & Sp(n_1) &  - & Sp(n_2) & - & 6{\bf F} \\
\end{array}
\ea\ee
where all internal links are again bifundamental half-hypermultiplets, have no consistent realizations.

Over all we thus find five consistent quivers of maximal depth for $(E_6,E_7)$ conformal matter
\be\ba\label{eq:QuiversE6E7}
&5{\bf F}-Sp(5)-Sp(5)-6{\bf F}\,,\\
&5{\bf F}-Sp(2)-SO(10)-Sp(3)-6{\bf F}\,,\\
&5{\bf F}-Sp(2)-SO(8)-Sp(4)-6{\bf F}\,,\\
&5{\bf F}-Sp(3)-SO(8)-Sp(3)-6{\bf F}\,,\\
&5{\bf F}-Sp(4)-SO(8)-Sp(2)-6{\bf F}\,.\\
\ea\ee

We repeat the analysis and determine consistent weakly couple quivers for  $(E_6,E_8)$ and $(E_7,E_8)$ conformal matter theories. Their flavor and gauge rank are
\be\ba\label{eq:GaugeFlavorConstraintE6E8}
(E_6,E_8)\,&:\quad \lb r_G,r_F\rb=\lb 18 ,15 \rb\,,\\
(E_7,E_8)\,&:\quad \lb r_G,r_F\rb=\lb 20 ,16 \rb\,,
\ea\ee
with the quivers of maximal depth of the structure
\be\ba\label{eq:possibleQuivers}
\tn{Maximal Depth Quivers}~(E_6,E_8)&:\qquad 5{\bf F}-Sp(n_1)-\prod G-Sp(n_2)-8{\bf F}\,,\\
\tn{Maximal Depth Quivers}~(E_7,E_8)&:\qquad 6{\bf F}-Sp(n_1)-\prod G-Sp(n_2)-8{\bf F}\,.
\ea\ee
These exhibit a manifest global symmetry of rank $15$ and $16$ respectively and consequently the interior must be empty. A hypermultiplet connecting the two $Sp$ gauge nodes would violate the constraint on the flavor rank and thus there are no potential quivers of type \eqref{eq:possibleQuivers} for $(E_6,E_8)$ and $(E_7,E_8)$ conformal matter.

We move on to consider quivers of submaximal depth for $(E_6,E_8)$ and $(E_7,E_8)$ conformal matter theories. These are of the structure
\be\ba\label{eq:SubMaxE6E8}
\tn{Submaximal Depth Quivers}~(E_6,E_8)&:\qquad 5{\bf F}-Sp(n_1)-\prod G-Sp(n_2)-5{\bf F}\,,\\
\tn{Submaximal Depth Quivers}~(E_7,E_8)&:\qquad 5{\bf F}-Sp(n_1)-\prod G-Sp(n_2)-6{\bf F}\,,
\ea\ee
as derived from the box graph embeddings in to the $(E_6,E_8)$ and $(E_7,E_8)$ CFDs in figures \ref{fig:CFDE6E8} and \ref{fig:CFDE7E8}. These quivers have a manifest flavor symmetry of rank $12,13$ whereby in both cases the interior $\prod G$ of the quiver is required to consist of either three gauge nodes connected by half-hypermultiplets or  two gauge nodes connected by a hypermultiplet. We analyse the case of $(E_6,E_8)$ in detail and state the result for $(E_7,E_8)$.

There are five choices of a connected interior $\prod G$ if we restrict to quivers without loops. The potential quivers for $(E_6,E_8)$ conformal matter theories of this kind read

\be\ba\label{eq:Quiver1}
5{\bf F}-Sp(n_1)-G_1-G_2-G_3-Sp(n_2)-5{\bf F}
\ea\ee
\vspace{5pt}
\renewcommand{\arraystretch}{0.9}
\setlength{\arraycolsep}{1.75pt}
\be\ba\label{eq:Quiver2}
&\begin{array}{ccccccccccc}
& & & & G_3 & & & &	& & \\
& & & & \bminus & & & & & & 	\\
	5{\bf F} & - & Sp(n_1) & - & G_1 &  - & G_2 & - & Sp(n_2) & - & 5{\bf F} \\
\end{array}
\ea\ee
\vspace{5pt}
\be\ba\label{eq:Quiver4}
&\begin{array}{ccccccccc}
& & & &G_3   & & & &	\\
& & & &\bminus  & & & &	\\
& & & &G_2   & & & &	\\
& & & &\bminus  & & & &	\\
	5{\bf F} & - & Sp(n_1) & - & G_1 & - & Sp(n_2) & - & 5{\bf F} \\
\end{array}
\ea\ee
\vspace{5pt}

\noindent where all interior links consist of half-hypermultiplets, as well as the quivers

\be\ba\label{eq:Quiver6}
5{\bf F}-Sp(n_1)-G_1-G_2-Sp(n_2)-5{\bf F}
\ea\ee
\vspace{5pt}
\be\ba\label{eq:Quiver5}
&\begin{array}{ccccccccc}
& & & &G_2   & & & &	\\
& & & &\bminus  & & & &	\\
	5{\bf F} & - & Sp(n_1) & - & G_1 & - & Sp(n_2) & - & 5{\bf F} \\
\end{array}
\ea\ee
\vspace{5pt}

\noindent where all interior links are given by half-hypermultiplets except for  $G_1-G_2$ links which consists of a full hypermultiplet. An analysis of the quiver type \eqref{eq:Quiver1} is very similar to that of maximal depth quivers and as only solutions we find 47 quivers of the structure\smallskip
\be\ba\label{eq:Quiver1E6E8}
5{\bf F}-Sp(n_1)-SO(r_1)-Sp(n_3)-SO(r_2)-Sp(n_2)-5{\bf F}\,.
\ea\ee\smallskip
\noindent A discussion of the quivers \eqref{eq:Quiver2}-\eqref{eq:Quiver6}, paralleling the following, shows that there are no consistent realizations of these. We therefore restrict the discussion to the final quiver \eqref{eq:Quiver5} of trinion topology. 

The gauge node $G_1$ of the quiver \eqref{eq:Quiver5} necessarily has a real, even dimensional fundamental representation and thus $G_1=F_4, SO(2r)$. The case $G_1=F_4$ is found to be inconsistent. For $G_1=SO(2r)$ the constraints from table \ref{tab:NumberConstraints} applied to the bottom line of the quiver \eqref{eq:Quiver5} together with the gauge rank constraint then read\smallskip
\be\ba\label{eq:ConstraintsR}
Sp(n_1)\,:&&\qquad\quad 5+\frac{2r}{2}&\leq 2n_1+6\,,\\
Sp(n_2)\,:&&\qquad\quad 5+\frac{2r}{2}&\leq 2n_2+6\,,\\
G_1\,:&&\qquad\quad \dim {\bf R}+\frac{2n_1}{2}+\frac{2n_2}{2}&\leq 2r-2\,,\\
r_{G}\,:&& n_1+n_2+r+\tn{rank}\,G_2&=18\,,
\ea\ee

\noindent where $\dim {\bf R}$ abbreviates the dimension of the lowest dimensional non-trivial representation of the undetermined gauge group $G_2$. We find $\dim {\bf R}\leq 8$ and applying the constraints from table \ref{tab:NumberConstraints} to the gauge node $G_2$ we find the only consistent choices to be $G_2=SU(n)_0$ with $n=5,6$. Here the subscript denotes a vanishing Chern-Simons level. The consistent quiver descriptions for $(E_6,E_8)$ conformal matter of quiver type \eqref{eq:Quiver5} are thus

\be\ba\label{eq:E6E8Quivers}
&\begin{array}{ccccccccc}
& & & &SU(6)_0   & & & &	\\
& & & &\bminus  & & & &	\\
	5{\bf F} & - & Sp(3) & - & SO(14) & - & Sp(3) & - & 5{\bf F} \\
\end{array} \\ ~ \\
&\begin{array}{ccccccccc}
& & & &SU(5)_0   & & & &	\\
& & & &\bminus  & & & &	\\
	5{\bf F} & - & Sp(3) & - & SO(14) & - & Sp(4) & - & 5{\bf F} \\
\end{array}
\ea\ee
\vspace*{5pt}

\noindent With a similar analysis we find the potential quiver of submaximal depth for $(E_7,E_8)$ conformal matter with connected interior to be of the structure
\be\ba\label{eq:E7E8Quiver1}
5{\bf F}-Sp(n_1)-SO(r_1)-Sp(n_3)-SO(r_2)-Sp(n_2)-6{\bf F}\,.
\ea\ee
The constraints on the number of hypermultiplets attaching to any one gauge node are solved by 45 integer tuples $(n_1,r_1,n_3,r_2,n_2)$. Here all internal links are bifundamental half-hypermultiplets.

\section{Conclusion and Outlook}

In this paper we reported on Calabi-Yau manifolds \eqref{eq:ResolutionSequenceMarginal}, \eqref{eq:blowups} realizing marginal 5d gauge theories in M-theory that originate from 6d $(E_n,E_m)$ conformal matter theories and derived their associated combined fiber diagrams which are given in figures \ref{fig:CFDE6E7}, \ref{fig:CFDE6E8}, \ref{fig:CFDE7E8}. These we used to constrain the list of quiver gauge theories of maximal and submaximal depth \eqref{eq:QuiversE6E7}, \eqref{eq:E6E8Quivers}, \eqref{eq:E7E8Quiver1} which in the strong coupling limit potentially complete to the respective SCFTs. 

The presented resolution of the $(E_n,E_m)$ singularities has no associated weakly coupled quiver gauge theory description as the geometries can not be consistently ruled. It would be interesting to study which quiver gauge theories can be realized by altering the resolution sequence such that the geometry allows for rulings. A systematic study of this requires understanding the structure with which the surface components $S_k$ glue to form the reducible surface $S=\cup_k S_k$. Describing which geometric transitions mediate between surfaces with different ruling would facilitate an enumeration of all quiver gauge theories associated to a marginal geometry and is one possible avenue for further research. For rank 2 theories this is achieved in \cite{Jefferson:2018irk}.

It is clear that the three constraints presented in section \ref{sec:Constraints} which restrict the potential quiver gauge theories are not sufficient. The list of quivers \eqref{eq:QuiversE6E7} for $(E_6,E_7)$ contain in part quivers which at the same level of reasoning are candidates to UV complete to descendants of $(E_7,E_7)$ conformal matter, cf. \cite{ Apruzzi:2019opn}. Further constraints generalising the results of \cite{Jefferson:2017ahm} such as recently explored in \cite{Bhardwaj:2020gyu} are needed to decide the UV behaviour of the proposed quivers.

The results of this paper complete the list of 5d theories originating from 6d conformal matter theories, as initiated in \cite{Apruzzi:2019vpe, Apruzzi:2019opn}, which are relevant to the discussion of circle reductions of the 6d UV-progenitor theories. The natural next step is the analysis of the higher rank progenitor theories with the final goal of systematizing all 5d theories that descend through Higgsable 6d SCFTs by RG flows induced through mass deformations and Higgs branch vacuum expectation values.

\subsection*{Acknowledgements}

It is a pleasure to thank Fabio Apruzzi, Marieke van Beest, Julius Eckhard, Sakura Sch\"afer-Nameki and Yinan Wang for helpful discussion and comments on the manuscript. The author is supported by the Studienstiftung des Deutschen Volkes.


\bibliography{5d}
\bibliographystyle{JHEP}

\end{document}